\documentclass[preprint,sort&compress]{elsarticle}

\usepackage[export]{adjustbox}[2011/08/13]

\usepackage{amsmath}
\usepackage{amsfonts}
\usepackage{amssymb}
\usepackage{siunitx}

\usepackage{cleveref}

\usepackage[colorinlistoftodos, color=green!40, prependcaption]{todonotes}
\setuptodonotes{inline}

\begin{document}

\begin{abstract}
Exploiting the framework of peridynamics, a dimensionally-reduced formulation for plates is developed that allows for the through-thickness nucleation and growth of fracture surfaces, enabling the treatment of delamination in a lower-dimensional model. Delamination fracture nucleation and propagation are treated by choosing the kinematics to be composed of an absolutely continuous part and a zone where jumps in the displacements are allowed. 
This assumption allows the explicit derivation of the dimensionally-reduced elastic energy, which shows a hierarchy of terms characterising the stored energy in a the plane element. 
An interpretation of the various terms of the reduced energy is shown by means of the simplest paradigm of bond-based peridynamics.
A striking feature of the reduced energy is that, despite the small-displacement assumption, there is a coupling between the membrane and bending terms. 
Semi-analytical solutions for simplified settings are obtained through a minimization procedure, and a range of nonstandard behaviors such as \textit{distal} crack nucleation and curved crack path are captured by the model.
Finally, the convergence of the proposed peridynamic reduced model to a local elastic theory for vanishing nonlocal lengthscale is determined, giving a local cohesive model for fracture.\\

\noindent Full article available at https://doi.org/10.1016/j.jmps.2022.105189.
\end{abstract}


\begin{frontmatter}
\title{Distal and non-symmetrical crack nucleation in delamination of plates via dimensionally-reduced peridynamics}
\author[1]{R. Cavuoto}
\author[1]{A. Cutolo}
\author[2,3]{K. Dayal}
\author[4,5,2,6,7]{L. Deseri\corref{cor1}}\ead{luca.deseri@unitn.it}
\author[1,8]{M. Fraldi\corref{cor1}}\ead{fraldi@unina.it}
\journal{Journal of the Mechanics and Physics of Solids}

\cortext[cor1]{Corresponding authors.}
\address[1]{Department of Structures for Engineering and Architecture, University of Naples, Italy}
\address[2]{Department of Civil and Environmental Engineering, Carnegie Mellon University, USA}
\address[3]{Center for Nonlinear Analysis, Department of Mathematical Sciences, CMU, USA}
\address[4]{Department of Civil, Environmental and Mechanical Engineering, University of Trento, Italy}
\address[5]{Department of Civil and Environmental Engineering, Pittsburgh University, PA, USA}
\address[6]{Department of Nanomedicine, Houston Methodist Hospital, Houston TX, USA}
\address[7]{Department of Mechanical Engineering, Carnegie Mellon University, Pittsburgh PA, USA}
\address[8]{Départment de Physique, Ecole Normale Supérieure, Paris, France}
\date{}
\end{frontmatter}

\section{Introduction}\label{ch:intro}

Delamination is a mode of failure that is typical of thin plate and shell structures in which the thickness is much smaller than the other two dimensions.
This mode of failure is characterised by a fracture in which the crack front propagates within the plane of the structure, resulting in the structure being broken up into layers. 
Composite laminates, which naturally present a weak plane at the interface between different materials, are especially vulnerable to delamination, but it can occur in microstructured and homogeneous thin structures as well \cite{INOUE2010, NAGANARAYANA1995}. 

This mode of fracture has been studied through a variety of approaches. 
Leading approaches include Cohesive Zone Models (CZM) \cite{DUGDALE1960,BARENBLATT1962,HILLERGORG1976,TURON2007,ELICES2002,FAN2008} and the extended finite element methods (XFEM) \cite{ZHAO2016,YAZDANI2016}. 
More recently, the phase-field technique for fracture has been used to model debonding in laminates \cite{KUMAR2021,ROY2017}. 
These models, however, treat the thin structure as a fully three-dimensional body and treat delamination explicitly.

In a distinct approach to the modeling of thin plates without accounting for delamination, an established procedure with a long history is to derive two-dimensional formulations for plates or films based on systematically reducing the three-dimensional theory using that the thickness is much smaller than the other dimensions.
This has been studied in a variety of settings, and are particularly attractive as they can lead to faster computational algorithms with good convergence properties while still capturing the key physical phenomena \cite{STEIGMANN2019,Arroyo-1,Arroyo-2,BAZILEVS2019,REDDY2019Geo,CONTI2021}. 

The aim of this work is to develop an approach that combines the advantages of dimensionally-reduced models while also accounting for delamination.
That is, we aim to derive a two-dimensional model of plates that allows for delamination failure.
Our approach is based on using peridynamics \cite{SILLING2000,KUNIN1975}, a nonlocal theory that models continuum bodies as a collection of infinitesimal material particles that interact through long-range forces, rather than the typical contact tractions.
In contrast to local theories of continuum mechanics that rely on the definition of strain, consequently constraining the displacement to have sufficient regularity, peridynamics works directly with the displacement and does not require regularity \textit{a priori}.
This makes it attractive to model damage, damage-fracture transition \cite{PRAKASH2022,DAYAL2021,DAYAL2022,CHUA2022,LIPTON2019c,DIEHL2022}, and dynamic phenomena such as impact and blasts \cite{JOOEUN2016,LIU2017,XU2008,ZHANG2018}.  

Papers dealing with two-dimensional peridynamic bodies can be divided into two main categories based on the approach: (1) full 3D numerical simulations \cite{BOBARU2011,LE2014,SAREGO2016}, and (2) 2D reduced models \cite{SILLING2005,TAYLOR2013,NAUMENKO2022,OGRADY2014,OGRADY20142,OGRADY20143,BAZILEVS2022,REDDY2016P,REDDY2013G,YANG2021}. While the former approach has been extensively used to treat delamination explicitly \cite{YOLUM2018,HU2015,XIAO2019}, existing reduced formulation only account for crack propagation with the crack tip oriented normal to the plane and thus cannot capture phenomena such as delamination.
In this work, a reduced formulation of \textit{bond-based} peridynamics, tailored to account for through-thickness delamination in thin plates characterized by a single material and no preexisting weak interface, is introduced.
As a first step, in Section \ref{ch:pdmodel}, the displacement field is additively decomposed into its absolutely continuous part and its jump to account for delamination fracture nucleation and propagation. 
That is, the natural function space for the displacement field is the space of functions of Special Bounded Variations (SBV), e.g. \cite{CHOKSI1999,GOBBINO1998}.
Further assumptions on both parts of the displacement field lead to a \textit{reduced form} of the peridynamics energy in Section \ref{ch:hierarchical}. 
The reduction procedure generates a hierarchy of terms characterising the strain energy stored inside the two-dimensional continuum element. 
A striking feature of the reduced energy is that, despite the small displacement assumption, there is a coupling between the membrane and bending terms. 
The hierarchy of the resulting functional allows for a consistent variational approach, enabling the displacement fields to be obtained by a minimization procedure.

Semi-analytical solutions for test cases are then obtained in Section \ref{ch:comparison}. 
The tests are performed on a thin cantilever plate, modeled with the proposed reduced formulation. 
In the first case, such a plate undergoes an imposed upward vertical displacement of the upper part of the free edge and a downward vertical displacement of the lower edge in a symmetrical manner - much like a peeling test. 
The model shows that variation of the nonlocal interaction lengthscale $\delta$, also called the horizon, induces different behaviors, namely distal or proximal damage nucleation. 
In the second case, an asymmetry is introduced by imposing the vertical upward displacement at various points of the upper edge of the plate, leading to non-symmetric crack propagation. 
In the third case, Mode-II fracture or sliding delamination is studied.
In order to explore the coupling between bending and membrane terms in the reduced formulation (which is geometrically linear) in a local setting, in Section \ref{ch:convergence}, we examine the convergence of the proposed model to a local theory when the nonlocal interaction scale $\delta$ tends to zero.
By enforcing a condition of bounded and non-vanishing energy, the scaling of the displacement field with $\delta$ is established; this, in turn, determines the scaling of all the terms in the energy, thereby allowing for the localization of the nonlocal model, leading to a reduced local formulation. 
The reduced local formulation has a cohesive structure, due to some terms of the energy associated with the jump part of the displacement surviving the limit operation.

\paragraph{Organization}
In Section \ref{ch:bondPD}, the constitutive framework of bond-based peridynamics is summarized. 
Section \ref{ch:pdmodel} sets up the theoretical framework for the reduced formulation, and the reduced elastic energy density of a peridynamic plate is obtained and interpreted.
The theoretical model is then implemented in a variational setting to be tested in simple loading conditions in Section \ref{ch:comparison}. 
Lastly, in Section \ref{ch:convergence}, the behavior of the reduced formulation proposed for vanishing horizon is investigated.

\section{Formulation of the peridynamics model} \label{ch:bondPD}
\begin{figure}[ht]
\footnotesize
\renewcommand{\figurename}{\footnotesize{Figure}}
	\begin{center}
	\includegraphics[width=0.49\textwidth]{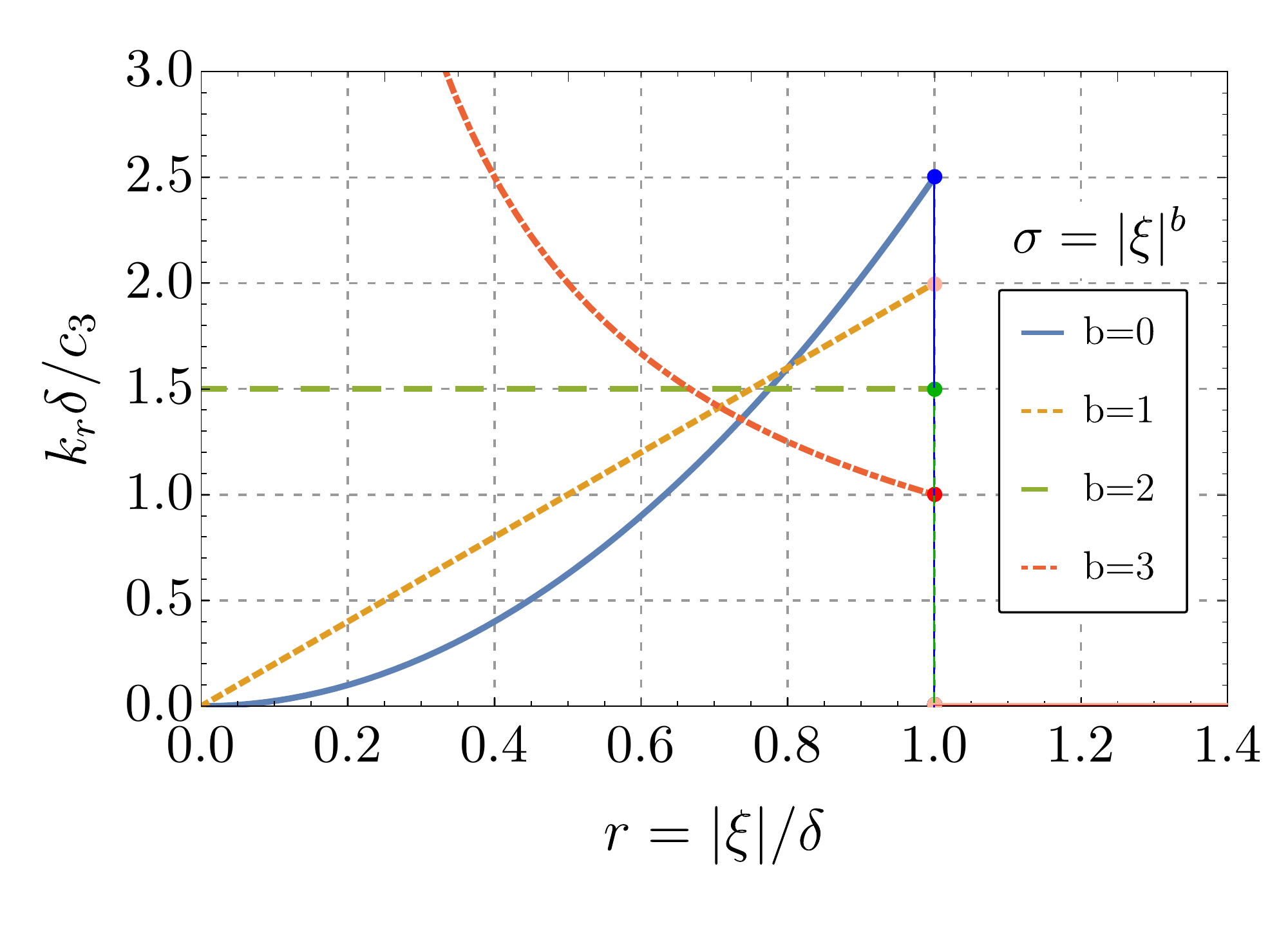}
	\includegraphics[width=0.49\textwidth]{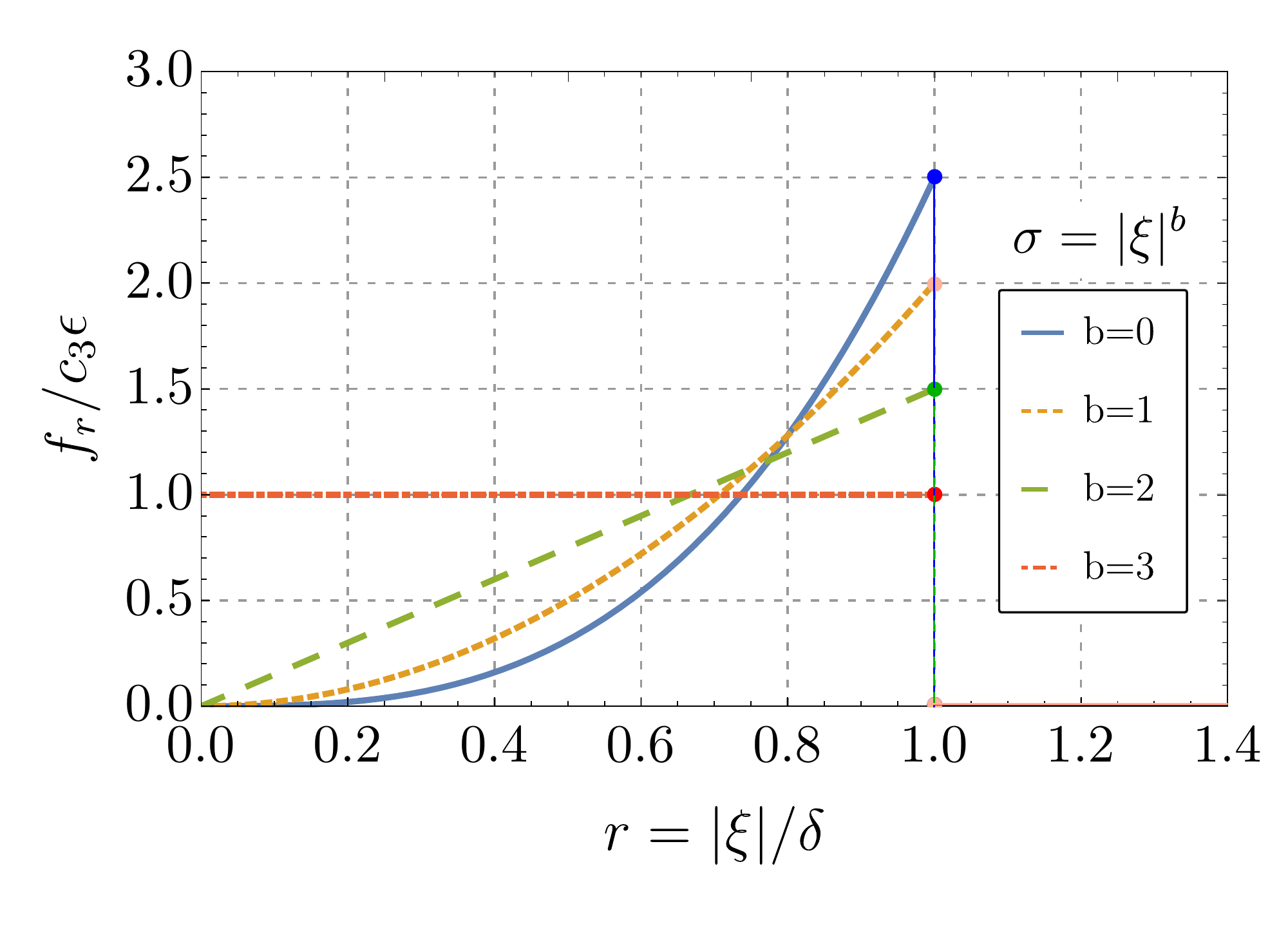}
	\includegraphics[width=0.49\textwidth]{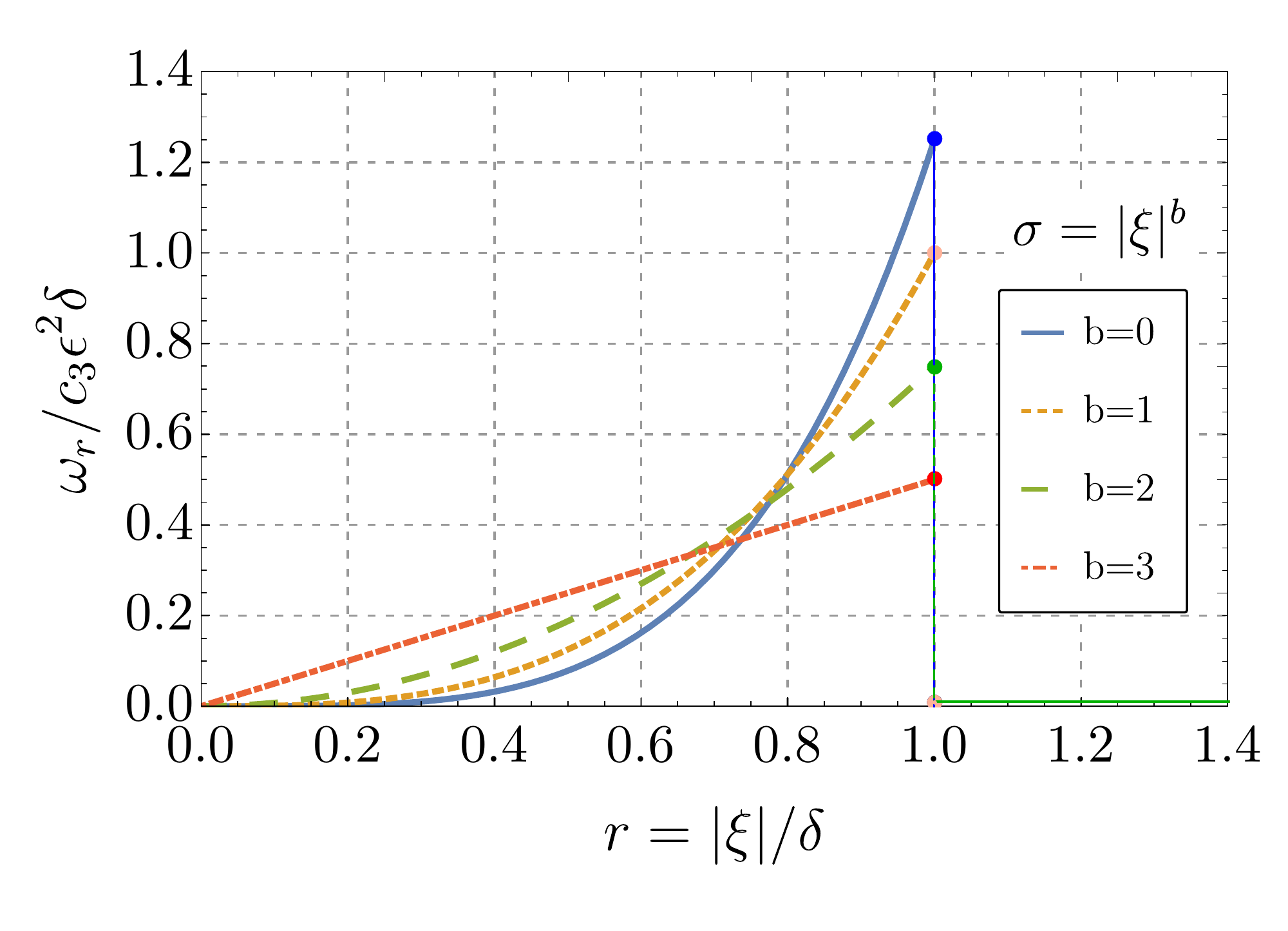}
	\end{center}
	\vspace{-10mm}
\caption{\footnotesize Upper left: normalized stiffness of a bond for different $\sigma$; $k_r=|\boldsymbol{f}|/|\boldsymbol{\eta}|$, $\delta$ is the horizon and $c_3$ the bond constant for the case of $\sigma=|\boldsymbol{\xi}|^3$ which recovers the original formulation of Silling \cite{SILLING2000}. Upper right: normalized bond force for different $\sigma$; $f_r$ is the modulus of the force as a function of $r$ (the normalized relative distance) that is applied to bonds under an imposed uniform stretch $\epsilon$. Below: the bond energy $\omega_r$ (energy per unit volume squared) for increasing bond length and imposed uniform stretch.}
\label{shapes}
\end{figure}

Bond-based peridynamics models a continuum body $\mathbb{B}$ as a collection of material particles interacting with one another, in pairs, through bonds. The equation for the static equilibrium of the body \cite{SILLING2000,DAYAL2006} can be written as the integrodifferential equation
\begin{equation}
    \int \limits _{\mathbb{H}} \boldsymbol{f} (\boldsymbol{x},\boldsymbol{x}',\boldsymbol{u},\boldsymbol{u}')\ \textup{d}V_{\boldsymbol{x}'}+\boldsymbol{b}(\boldsymbol{x})=0\ ,
\label{eq:motion}
\end{equation}
where: $\boldsymbol{f}$, called the pairwise force field, is the force exerted between material particle $\boldsymbol{x}$ and $\boldsymbol{x}'$, it can depend on the displacements $\boldsymbol{u}$ of such particles and hosts all the constitutive information of the model; $\boldsymbol{b}$ is the vector of external body force; $\mathbb{H}$ represents the so-called \textit{family} of $\boldsymbol{x}$ and is the set of all the material points that are within a characteristic distance from it, called horizon, $\delta$. In bond-based peridynamics, the balances of linear and angular momentum require the pairwise force field, $\boldsymbol{f}$, to be anti-symmetric with respect to particles switch \cite{SILLING2000}
\begin{equation}
\boldsymbol{f}(\boldsymbol{x},\boldsymbol{x}',\boldsymbol{u},\boldsymbol{u}')=-\boldsymbol{f}(\boldsymbol{x}',\boldsymbol{x},\boldsymbol{u}',\boldsymbol{u})\ .
\label{eq:LAC}
\end{equation}
Constitutive relations for the definition of $\boldsymbol{f}$ have been proposed by many authors \cite{SILLING2000,DU2011}, among which one of the simplest is the standard linear elastic perfectly brittle relation revisited by Zhou \cite{ZHOU2010},
\begin{equation}
\boldsymbol{f}=\mu\ c\ s\ \frac{|\boldsymbol{\xi}|^2}{\sigma}\boldsymbol{\xi}\ ,
\label{eq:emmirch}
\end{equation}
where $c$ is called the \textit{bond constant} (a positive scalar quantity), $\boldsymbol{\xi}$ and $\boldsymbol{\eta}$ are the relative position and displacement of the particles respectively, and $s$ is the stretch of the bond. Lastly, $\sigma=\sigma(\boldsymbol{\xi})$ is a function ensuring integrability of (\ref{eq:emmirch}). 
Additional conditions on $\sigma(\boldsymbol{\xi})$ are necessary to bound the stiffness and the energy respectively of the PD model to finite positive values. 
The effect of $\sigma(\boldsymbol{\xi})$ on the PD model, equation (\ref{eq:emmirch}), is depicted in Figure \ref{shapes}. The function $\mu=\mu(\boldsymbol{\xi},\boldsymbol{\eta})$ in equation (\ref{eq:emmirch}) is a history-dependent scalar-valued function (also called \textit{failure parameter}) which enforces bond breakage under tension only:
\begin{equation}
\mu(\boldsymbol{\xi},\boldsymbol{\eta})= \left\{
\begin{tabular}{l l}
$1$&\textup{for} $s<s_{\textup{cr}}$\\
$0$&\textup{otherwise}
\end{tabular}
 \right. ,
\label{eq:mi}
\end{equation}
where $s_{cr}$ is a critical threshold for the bond elongation. Form (\ref{eq:emmirch}) admits a potential, called \textit{pairwise potential function}:
\begin{equation}
\omega(\boldsymbol{\xi},\boldsymbol{\eta})=\int \boldsymbol{f}(\boldsymbol{\xi},\boldsymbol{\eta})\cdot \textup{d}\boldsymbol{\eta}=
\left\{
\begin{tabular}{l l}
$\omega_{el}=\frac{1}{2}c\ s^2\frac{|\boldsymbol{\xi}|^4}{\sigma}$&\textup{for} $s<s_{\textup{cr}}$\\
$\omega_{\textup{cr}}=\frac{1}{2}c\ s_{cr}^2\frac{|\boldsymbol{\xi}|^4}{\sigma}$&\textup{otherwise}
\end{tabular}
 \right. .
\label{eq:energyPD}
\end{equation}

\subsection{A microstructural interpretation of peridynamics}
\label{ch:motivations}
\begin{figure}
\footnotesize
\renewcommand{\figurename}{\footnotesize{Figure}}
\centering
\begin{minipage}{.5\textwidth}
  \centering
  \includegraphics[width=\linewidth]{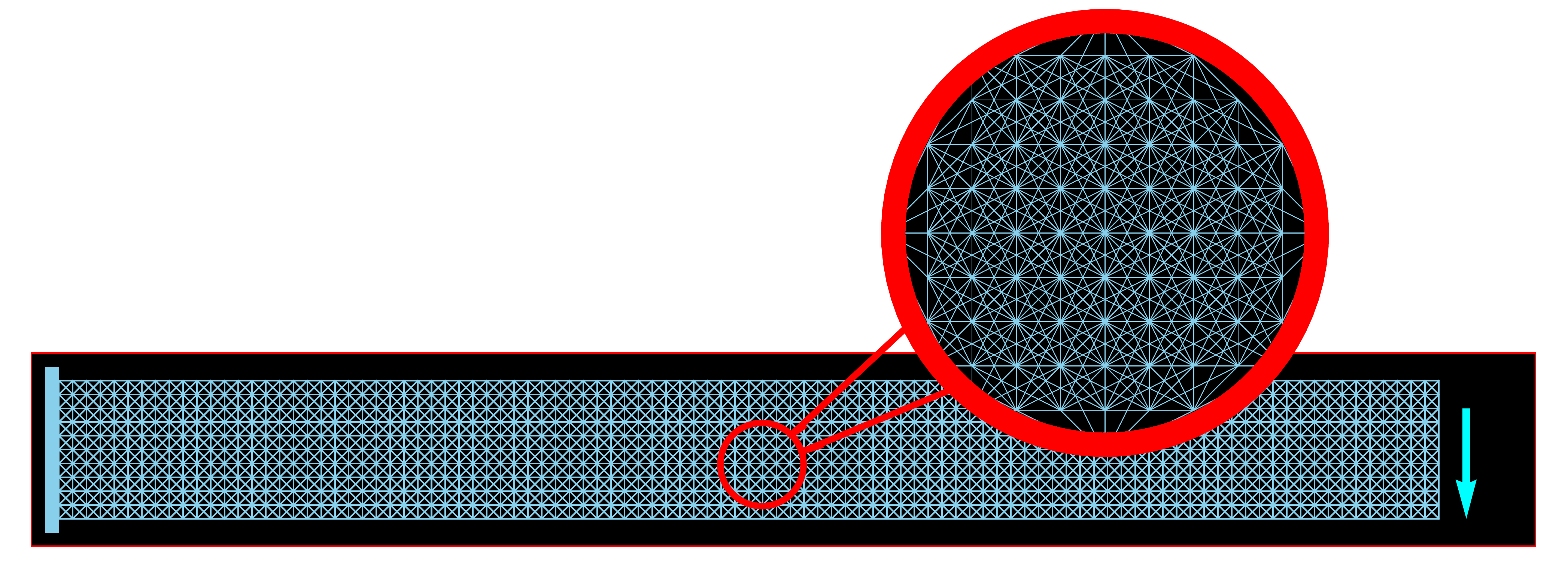}
\end{minipage}%
\begin{minipage}{.5\textwidth}
  \centering
  \includegraphics[width=\linewidth]{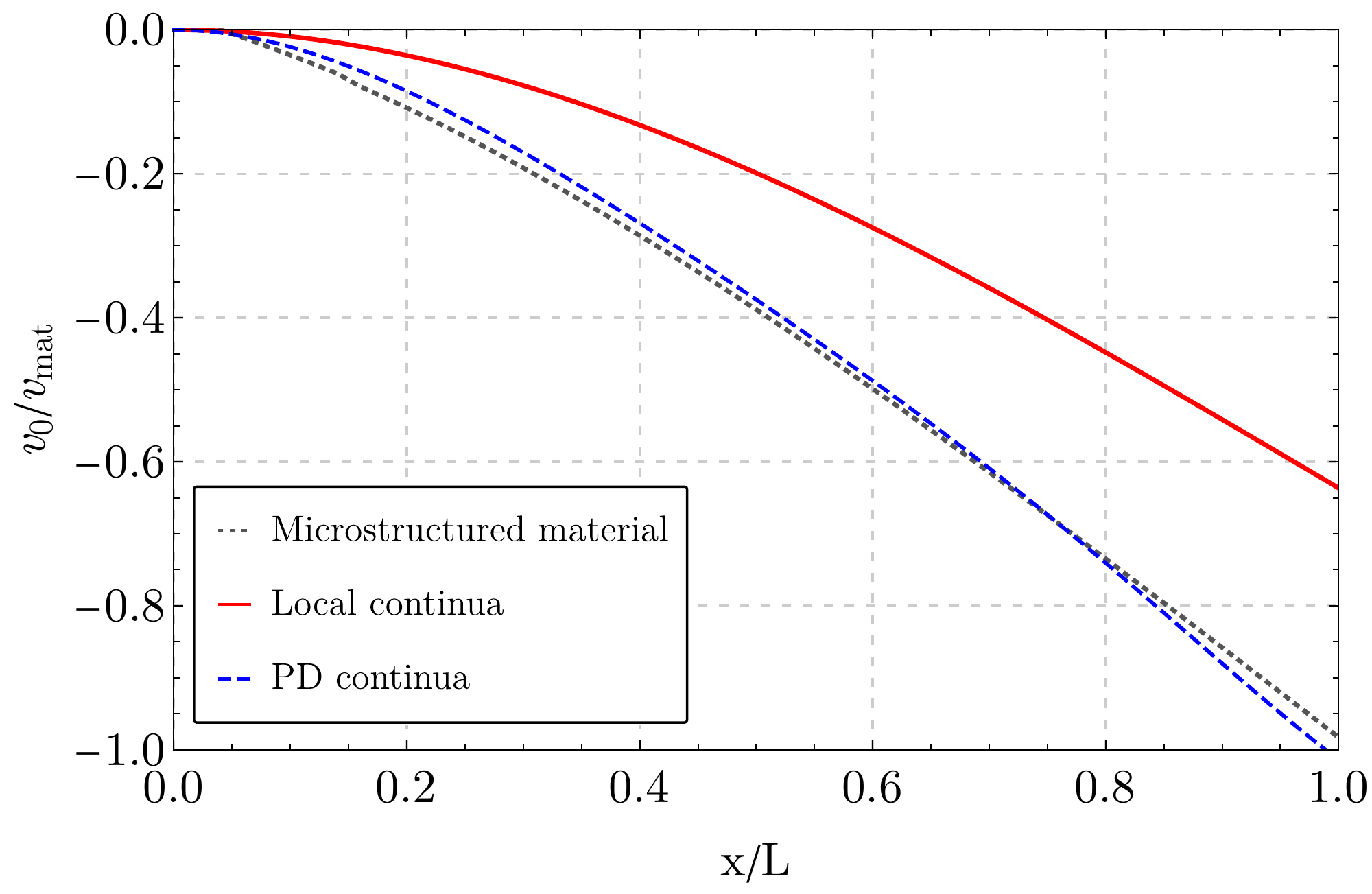}
\end{minipage}
\caption{\footnotesize (Left) Cantilever beam characterized by a net-like micro-structure and subjected to an imposed displacement at its free end. (Right) The normalized deformed shapes ($v_{\textup{mat}}$ maximum vertical displacement of structured material) according to peridynamic (PD) and local theory.}
\label{fig:microresponse}
\end{figure}

The present study proposes a dimensionally reduced formulation of peridynamic plates with a particular focus on through-thickness fracture propagation. Since the model, developed in Section \ref{ch:pdmodel}, shows unconventional scalings of the energy terms due to its nonlocal (peridynamic) nature, in the present preliminary section a possible micro-structural interpretation of peridynamics, functional to the mechanical characterization of our model, is discussed.\\
The possibility to pass from the continuum to the discrete level is crucial in problems involving the transition from elastic to dissipative phenomena such as damage and fracture. Nevertheless, the exact equivalence between a given peridynamic model and a corresponding microstructure is usually not trivial to find. One possible interpretation, available from purely energetic arguments, can be given by means of a discrete structure, see \ref{appendix2}. To stress such observations, we build a simple numerical example in which a structure made of interconnected linear elastic elements leads to a densely packed truss ensemble. Despite one would assume that the asymptotic behavior for an increasing number of micro-beams of the structure tends to that of a standard local continuum, we demonstrate that --for a prescribed topology-- significant discrepancies in terms of displacements emerges between discrete and homogenised local continuum.
Figure \ref{fig:microresponse} depicts the case of a cantilever beam (1 meter long and 0.4 meters thick), built by assembling trusses in a net-like pattern as displayed in the inset of such figure. These trusses are connecting each material point of the body with all the others that satisfy a relative distance requirement. The parameters involved, namely axial stiffness of the beams and horizon length, are calibrated in such a way that the behavior under tension reproduces that of an ideal homogenized continuous local beam. In Figure \ref{fig:microresponse} the deflection of the structured beam when subjected to a vertical force applied at one end is compared with local theories (in red Euler-Bernoulli and Timoshenko beam overlapping one onto the other) and with the predictions of peridynamics (PD, in blue). Failing of local theories to correctly characterize the bending behavior for the example introduced above, can be ascribed to the intricate internal structure of the beam.

\section{A dimensionally-reduced model for thin plates} \label{ch:pdmodel}

In order to develop the analytical calculations necessary for the formulation of a dimensionally reduced model for plates, certain assumptions are made on both the kinematics of the plate and on the constitutive relation of the bond-based peridynamic continua.

\subsection{Kinematics}

Since the fracture of a material can be seen as the nucleation and growth of a discontinuity in its displacement field, one can additively partition the kinematics into a continuous part, accounting for elastic deformations, and a jump part, accounting for the displacements due to the delamination, namely:
\begin{equation}
    \boldsymbol{u}(\boldsymbol{x})=\boldsymbol{u}_{a}(\boldsymbol{x}) +\boldsymbol{u}_{J}(\boldsymbol{x})\ ,
\label{eq:totdisp}
\end{equation}
where the index $a$ indicates the absolutely continuous part while the index $J$ denotes the jump part. In this sense, it can be said that $\boldsymbol{u}$ is a function in the space of Special Bounded Variations (SBV) \cite{GOBBINO1998,CHOKSI1999}.

\begin{figure}[ht]
\footnotesize
\renewcommand{\figurename}{\footnotesize{Figure}}
	\begin{center}
	\includegraphics[width=\textwidth,angle=0]{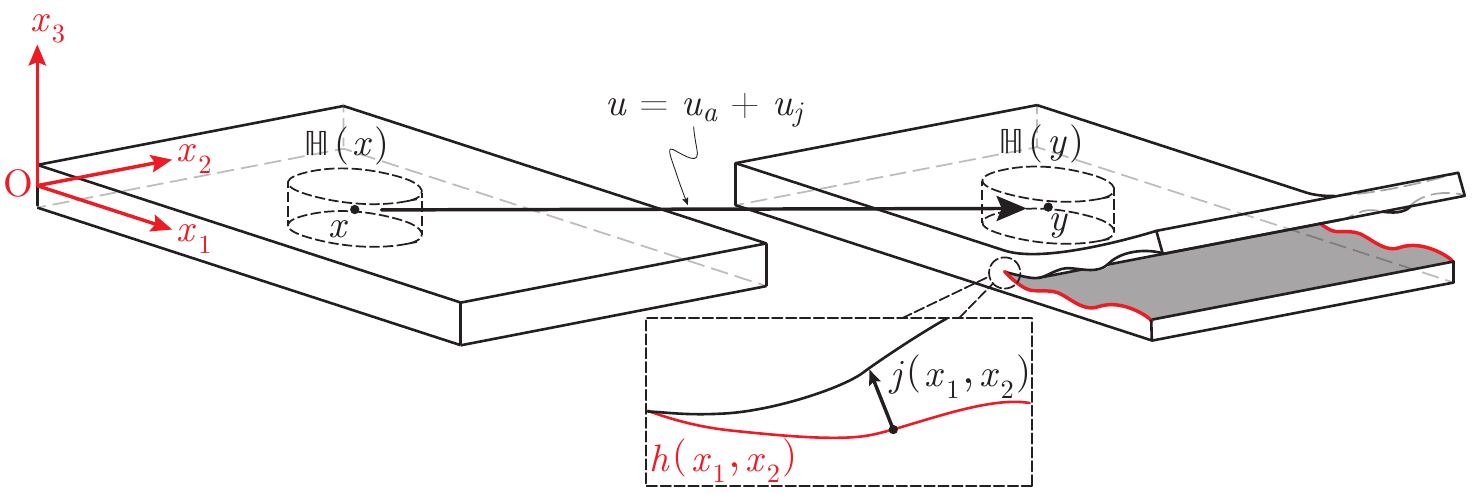}
	\end{center}
	\vspace{-5mm}
\caption{\footnotesize Kinematics of a plate of thickness H undergoing through-thickness fracture propagation, here also referred as delamination. Each point of the reference configuration lying on possible delamination surfaces to be determined jumps to a new position specified by the vector $\boldsymbol{j}(x_1,x_2)$.}
\label{reference}
\end{figure}

\noindent We now restrict ourselves to the study of thin bodies, $\mathbb{B}$, characterized by a constant thickness H. Given a region of the three-dimensional euclidean space $\mathbb{E}^3$, and a Cartesian reference frame $(O,x_1,x_2,x_3)$, Figure \ref{reference}, the absolutely continuous part of the displacement is approximated by a polynomial expansion as follows
 \begin{equation}
\boldsymbol{u}_{a}(\boldsymbol{x})\approx \boldsymbol{\mathcal{A}}(x_1,x_2)+\boldsymbol{\mathcal{B}}(x_1,x_2)x_3+\cdots \ \ .
\label{continuousfield}
\end{equation}
It is important to highlight that the choice of the reduction plane (the expansion point in the expansion above) can have effects on the hierarchical distribution of terms in the reduced formulation \cite{TAYLOR2013}. In the sequel the midplane of the plate is chosen to perform the dimension reduction, so as to align with classical local elastic reduced formulations.\\

\noindent The jump part of the displacement field will be represented in a rather general way as
\begin{equation}
\boldsymbol{u}_{J}(\boldsymbol{x})=\boldsymbol{j}(x_1,x_2)\cdot  \Theta(x_3-h(x_1,x_2)),
\label{eq:jumpdispl}
\end{equation}
where the $h(x_1,x_2)$ can be regarded as the crack surface, defining the surface on which a displacement discontinuity may arise, while $\Theta$ is the Heaviside function; lastly, $\boldsymbol{j}(x_1,x_2)$ is the vector function defining the jump itself. \textit{It is worth noting that mixed-mode fracture processes are allowed by the ansatz made above on} $\boldsymbol{u}_J$. Due to the kinematic split imposed on equation (\ref{eq:totdisp}), the relative displacement field now reads as follows:
\begin{equation}
\boldsymbol{\eta}=\boldsymbol{\eta}_{a}+\boldsymbol{\eta}_{J}.
\label{eq:reldisp}
\end{equation}

\subsection{Damage} \label{damage}

\begin{figure}[ht]
\footnotesize
\renewcommand{\figurename}{\footnotesize{Figure}}
	\begin{center}
	\includegraphics[width=0.6\textwidth,angle=0,trim={0cm 0cm 0cm 0cm},clip]{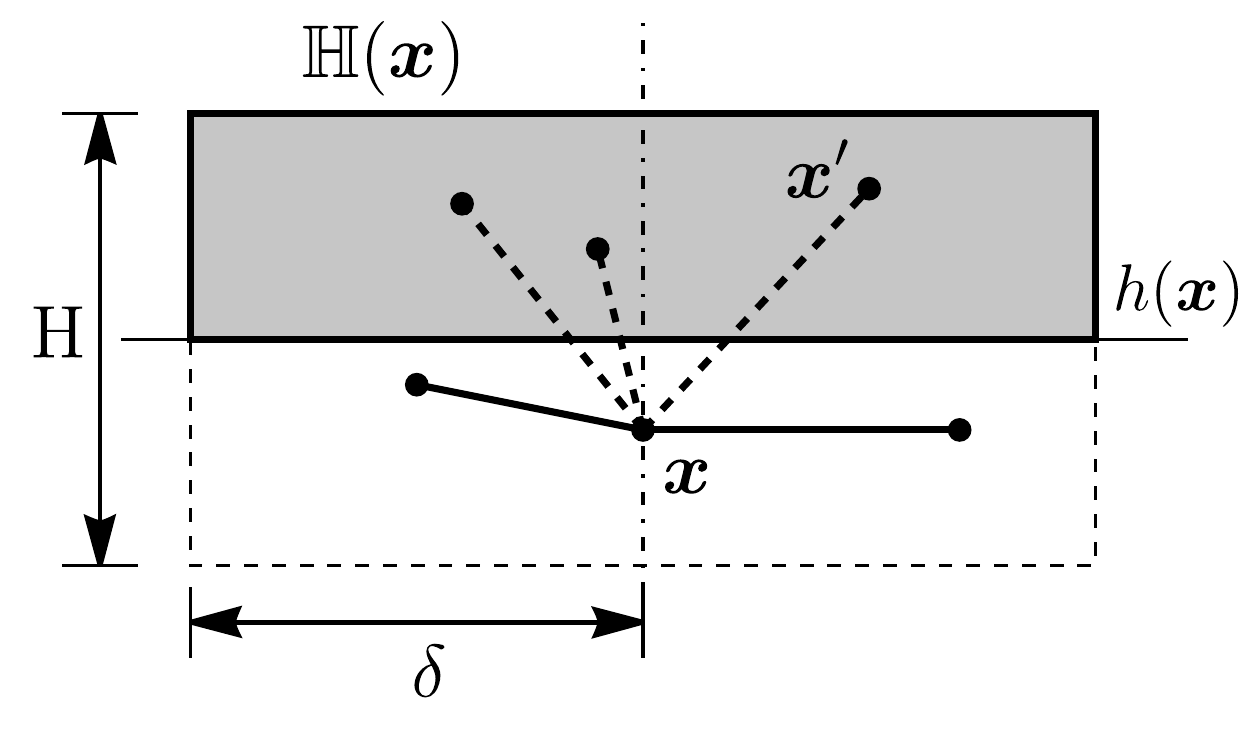}
	\end{center}
	\vspace{-5mm}
\caption{\footnotesize Computation of the total energy necessary to break all the bonds connecting the material point $\boldsymbol{x}$ with those $\boldsymbol{x'}$ on the other side of the fracture surface $h(\boldsymbol{x})$.}
\label{fracture_surface}
\end{figure}

It is important to highlight that in nonlocal theories a discontinuity in the displacement field does not necessarily mean fracture nucleation/propagation, as particles that are already separated by a finite distance can very well withstand a jump in their relative displacement. In PD, what ensures the effective occurrence of damage is the $\mu$ function (\ref{eq:mi}), which represents the failure criterion for the bonds. 
Indeed, the state of interaction can be determined by means of equation (\ref{eq:mi}), which enforces a critical stretch condition ($s<s_{\textup{cr}}$) \cite{SILLING2000,ERDOGAN2008,BOBARU2009}. In many other cases available in the literature, instead of a critical elongation criterion, an energy-based one is employed \cite{FOSTER2011,ZHANG2020,MADENCI2016}. Such a criterion relates the breakage of a bond to the attainment of a threshold in the stored energy, called critical bond energy $\omega_{\textup{cr}}$. Both the critical stretch and critical energy are typically evaluated by means of an energy comparison with the standard local theory of fracture mechanics.
In particular, the PD energy necessary for the growth of a new surface in the body, defined as the energy required to break all the bonds which pass through that particular surface (Figure \ref{fracture_surface}), is imposed to be equal to the critical energy release rate of Griffith theory \cite{GRIFFITH1921}, an operation that ensures the recovery of the Griffith theory in the limit of small horizon \cite{LIPTON2021f,LIPTON2019c,JHA2018}.

In this fashion, for the case of the critical stretch, one obtains \cite{SILLING2005,LEHOUCQ2010} 
\begin{equation}
s_{cr}=\sqrt \frac{G_c}{\beta (\mathbb{H},\sigma)\ \delta}\ ,
\label{griffith}
\end{equation}
where $G_c$ is the critical energy release rate and $\beta$ is a scalar function of the shape of the family $\mathbb{H}$ and on $\sigma(|\boldsymbol{\xi}|)$.\\
The transition from damage to fracture is therefore naturally tracked by the failure mechanism of PD. 

\subsection{Lagrangian formulation} \label{ch:lagrangian}
Under certain conditions \cite{breitzman2018bond}, the solution of the equilibrium problem of the nonlocal PD body coincides with the stationary points of the following functional \cite{MENGESHA2015,BELLIDO2014,AKSOYLU2010}: 
\begin{equation}
\mathcal{L}=-\mathcal{E}_{el}+\mathcal{W},
\label{eq:lagrangian}
\end{equation}
where $\mathcal{E}_{el}$ is the elastic energy, and $\mathcal{W}$ is the work of the external loads. By explicitly expressing the various terms, equation (\ref{eq:lagrangian}) becomes
\begin{equation}
\mathcal{L}[\boldsymbol{u}]=-\frac{1}{2}\int \limits _{\mathbb{B}} \int \limits _{\mathbb{H}}\omega(\boldsymbol{\xi},\boldsymbol{\eta})\  \textup{dV}'\textup{dV}
+\int \limits_{\mathbb{B}} \boldsymbol{b}\cdot \boldsymbol{u}\ \textup{dV}\ ,
\label{eq:lagrangian2}
\end{equation}
\noindent where $\omega$ is the energy density defined in (\ref{eq:energyPD}), while $\mathbb{B}$ and $\mathbb{H}$ are the continuum body and the family of a point, respectively.
We here define the decomposition of $\mathbb{B}$ through a Cartesian product as $\mathbb{B}=\mathbb{B}_{\alpha}\times \mathbb{B}_3$, where 
\begin{equation*}
\mathbb{B}_{\alpha}:=\left\{(x_1,x_2),\ \forall \boldsymbol{x}\in \mathbb{B}\right\}\ \ \textup{and} \ \ \mathbb{B}_{3}:=\left\{x_3,\ \forall \boldsymbol{x}\in \mathbb{B}\right\}.
\label{eq:domainsB}
\end{equation*}
Accordingly, one can define $\mathbb{H}=\mathbb{H}_{\alpha}\times \mathbb{H}_3$, where 
\begin{align*}
\mathbb{H}_{\alpha}&:=\left\{(x'_1,x'_2),\ \forall \boldsymbol{x}'\in \mathbb{B}: \sqrt{(x'_1-x_1)^2+(x'_2-x_2)^2} \leq \delta\right\}\ , \\
\mathbb{H}_{3}&:=\left\{x'_3,\ \forall \boldsymbol{x}'\in \mathbb{B}: \sqrt{(x'_1-x_1)^2+(x'_2-x_2)^2} \leq \delta\right\}.
\label{eq:domainsH}
\end{align*}
Since in the present study $x_3$ is chosen as the out-of-plane coordinate (see Figure \ref{reference}), its value ranges in between $\{-\textup{H}/2,\textup{H}/2\}$, H being the plate thickness.\\
In view of the previous Cartesian products, one can now write
\begin{equation}
\mathcal{L}[\boldsymbol{u}]=\int \limits_{\mathbb{B}_{\alpha}}\left(-\frac{1}{2}\int \limits _{\mathbb{H}_{\alpha}}\int \limits _{\mathbb{H}_3} \int \limits _{\mathbb{B}_3}\omega(\boldsymbol{\xi},\boldsymbol{\eta})\  \textup{d}x_3\textup{d}x_3'\textup{dS}_{\alpha}+\int \limits _{\mathbb{B}_3}\boldsymbol{b}\cdot \boldsymbol{u}\ \textup{d}x_3\right)\textup{dS}_{\alpha}.
\label{eq:lagrangianexpl}
\end{equation}
\noindent Performing the integrations through the thickness of (\ref{eq:lagrangianexpl}) allows one to obtain the reduced form of the total Lagrangian of the plate. In particular, the first addend in parenthesis of equation (\ref{eq:lagrangianexpl}), which is the elastic energy per unit surface $\Lambda_{\mathcal{E}}$, becomes: 
\begin{align}
\Lambda_{\mathcal{E}}&=\frac{1}{2}\int \limits _{\mathbb{H}_{\alpha}}\int \limits _{\mathbb{H}_3} \int \limits _{\mathbb{B}_3}\omega(\boldsymbol{\xi},\boldsymbol{\eta})\  \textup{d}x_3\textup{d}x_3'\textup{dS}_{\alpha}= \nonumber \\
&=\frac{1}{2}\int \limits _{\mathbb{H}_{\alpha}}\int \limits _{-\frac{\textup{H}}{2}}^{\frac{\textup{H}}{2}}\int \limits _{-\frac{\textup{H}}{2}}^{\frac{\textup{H}}{2}} \omega(\boldsymbol{\xi},\boldsymbol{\eta})\textup{d}x_3\textup{d}x_3'\textup{dS}_{\alpha}=\int \limits _{\mathbb{H}_{\alpha}}\omega_{\textup{red}}\textup{dS}_{\alpha}\ ,
\label{eq:ered1}
\end{align}
where $\omega_{\textup{red}}$ is the reduced form of the pairwise potential function $\omega$.\\
Similarly, we refer to the result of the through-thickness integration of the work of the external loads (second addend in parenthesis in equation (\ref{eq:lagrangianexpl})) as $\Lambda_{\mathcal{W}}$.\\
\noindent All the functionals involved in (\ref{eq:lagrangianexpl}) are nonlocal, as the unknown function $\boldsymbol{u}$ is evaluated at different points of the body. An equivalent form of the Euler-Lagrange equation for nonlocal functionals is now necessary to find the stationary points of (\ref{eq:lagrangianexpl}). The search for stationary points within the interior of the domain of the functional (or its minimization) has been investigated in \cite{FOSS2018}. For the particular case of static and elastic PD nonlocal functional \cite{MENGESHA2015,BELLIDO2014,AKSOYLU2010} one has that the following implication holds:
\begin{equation}
\min \limits_{u} \mathcal{L} \rightarrow 2\frac{\partial \Lambda_{\mathcal{E}}}{\partial \boldsymbol{q}}-\frac{\partial \Lambda_{\mathcal{W}}}{\partial \boldsymbol{q}}=0\ ,
\label{ELeqs}
\end{equation}
where the $\boldsymbol{q}$ is the vector of the unknown functions of the problem, which because of \cref{eq:totdisp,continuousfield,eq:jumpdispl} reads as follows: $$\boldsymbol{q}=\{h(x_1,x_2),\boldsymbol{j}(x_1,x_2),\boldsymbol{\mathcal{A}}(x_1,x_2),\boldsymbol{\mathcal{B}}(x_1,x_2)\}\ .$$
\noindent  In order to retrieve equation (\ref{ELeqs}), condition (\ref{eq:LAC}) must be enforced on the results of \cite{MENGESHA2015,BELLIDO2014,AKSOYLU2010}.

\subsection{Hierarchical form of the reduced pairwise potential function} \label{ch:hierarchical}
 We here retrieve an explicit form of the reduced pairwise potential function,
\begin{equation}
\omega_{\textup{red}}=\int\limits _{-\frac{\textup{H}}{2}}^{\frac{\textup{H}}{2}}  \int\limits _{-\frac{\textup{\textup{H}}}{2}}^{\frac{\textup{H}}{2}}   \omega(\boldsymbol{\xi},\boldsymbol{\eta})\ \textup{d}x_3\textup{d}x_3'\ ,
\label{eq:ered}
\end{equation}
for a bond-based peridynamic body. For the linear elastic case, the influence of the function $\sigma$ appearing in (\ref{eq:emmirch}) on the overall behavior has been indirectly investigated in Bobaru at al. \cite{BOBARU2009}. There, the authors have shown how the shape of the micromodulus function has indeed consequences on the overall behavior of the material, albeit this does not influence the rate of convergence for a vanishing horizon. \textit{The result is critical to this work}, where the consequences of a localization procedure on the reduced peridynamic model will be explored (section \ref{ch:convergence}) with the objective of retrieving a local reduced formulation for plates.\\
For the purpose of simplifying the calculations, drawing on the results discussed above \cite{BOBARU2009}, condition $\sigma=1$ is enforced in the sequel. Neglecting the failure parameter (denoted by $\mu$ in equation (\ref{eq:mi})) allows the evaluation of the reduced form of the energy for the fully elastic case, i.e. when the load has yet to break any bond. If $c$ is the bond constant, and $\phi$ is the ratio between the in-plane component of the horizon and the thickness, then
\begin{equation}
    \frac{\omega_{\textup{red}}}{c}=\underbrace{\textup{H}^2\ p_1(\phi,\boldsymbol{u}_a)+\textup{H}^4\ p_2(\phi,\boldsymbol{u}_a)+\textup{H}^6\ p_3(\phi,\boldsymbol{u}_a)}_{%
    \let\scriptstyle\textstyle
    \substack{\omega_{\mathrm{red},a}}}  + \omega_{\textup{red,J}}(\textup{H},\phi,\boldsymbol{u}_a,\boldsymbol{u}_J)\ \ , \label{eq:wred}
\end{equation}
where
\begin{flalign*}
    p_1(\phi,\boldsymbol{u}_a)=&(2\phi-\phi^2)(x_1'-x_1)^2 \left(\mathcal{A}_1(x_1')-\mathcal{A}_1(x_1)\right)^{2}/4\ \ ;\\ 
    p_2(\phi,\boldsymbol{u}_a)=& \left\{ \right. 2\phi^3(3\phi-4)(\mathcal{A}_3(x_1')-\mathcal{A}_3(x_1))^2+\\ 
& (x_1'-x_1)^2\phi^3(3\phi-4)(\mathcal{B}_1(x_1')^2+\mathcal{B}_1(x_1)^2)+\\
& (x_1'-x_1)^2(-2\phi+3\phi^2-\phi^4)(\mathcal{B}_1(x_1')-\mathcal{B}_1(x_1))^2 +\\
& 2\phi^3(3\phi-4)(x_1'-x_1)(\mathcal{A}_1(x_1')-\mathcal{A}_1(x_1))(\mathcal{B}_3(x_1')+\mathcal{B}_3(x_1))+\\
& \left. 2\phi^3(3\phi-4)(x_1'-x_1)(\mathcal{A}_3(x_1')-\mathcal{A}_3(x_1))(\mathcal{B}_1(x_1')+\mathcal{B}_1(x_1)) \right\} /48\ \ ;\\
p_3(\phi,\boldsymbol{u}_a)=& \{\left(20-45\phi+72\phi^2-80\phi^3 \right)\left(\mathcal{B}_3(x_1')-\mathcal{B}_3(x_1) \right)^2 +  \\ 
& \left. (20-45\phi-72\phi^2+80\phi^3)(\mathcal{B}_3(x_1')\mathcal{B}_3(x_1))\right\}/1440 \ \ ;
\end{flalign*}
\noindent while 
$\omega_{\textup{red,J}}$, an implicit function of $\phi$, H and the unknown fields, denotes the part of the reduced energy associated with the jump field. As shown in equation (\ref{eq:wred}) the part of the reduced energy associated with the continuous displacements is henceforth denoted by $\omega_{\textup{red,a}}$. \\
In the case of a through-thickness horizon equal to the whole thickness of the thin element, the physical condition of isotropic interaction is assumed. In such a case:
\begin{align}
    p_1(1,\boldsymbol{u})=& (x_1'-x_1)^2 \left(\mathcal{A}_1(x_1')-\mathcal{A}_1(x_1)\right)^2/4\ \ ;\label{eq:p1}\\ 
    p_2(1,\boldsymbol{u})=& \left\{ \right. -2(x_1'-x_1)(\mathcal{A}_3(x_1')-\mathcal{A}_3(x_1))(\mathcal{B}_1(x_1')+\mathcal{B}_1(x_1))+ \label{eq:p2}\\
& -2(\mathcal{A}_3(x_1')-\mathcal{A}_3(x_1))^2-(x_1'-x_1)^2(\mathcal{B}_1(x_1')^2+\mathcal{B}_1(x_1)^2)+\nonumber \\
& -2(x_1'-x_1)(\mathcal{A}_1(x_1')-\mathcal{A}_1(x_1))(\mathcal{B}_3(x_1')+\mathcal{B}_3(x_1)) \} /48\ \ ;\nonumber \\
p_3(1,\boldsymbol{u})=& \{ 10\mathcal{B}_3(x_1') \mathcal{B}_3(x_1)+7 \mathcal{B}_3(x_1'){}^2+7 \mathcal{B}_3(x_1){}^2\} /1440 \ \ ; \label{eq:p3}
\end{align}
and:
\begin{align}
    \frac{\omega_{\textup{red},J}}{c}=&\ r_0(\boldsymbol{u}_J)+\textup{H}\ r_1(\boldsymbol{u}_a,\boldsymbol{u}_J)+\textup{H}^2\ r_2(\boldsymbol{u}_J)+\\
    &\ \textup{H}^3\ r_3(\boldsymbol{u}_a,\boldsymbol{u}_J)+\textup{H}^4\ r_4(\boldsymbol{u}_a,\boldsymbol{u}_J)+\textup{H}^5\ r_5(\boldsymbol{u}_a,\boldsymbol{u}_J) \nonumber\ \ \ ,
\end{align}
where the $r_i$ functions are reported in \ref{ch:appendix1}.

For simplicity, equation (\ref{eq:wred}) has been specialized for the plane strain case. The variables $x_1$ and $x_1'$ represent respectively the in-plane component of the position vector for particle $\boldsymbol{x}$ and $\boldsymbol{x}'$. Furthermore, we have used $\boldsymbol{\mathcal{A}}=\{\mathcal{A}_1,\mathcal{A}_3\}$, $\boldsymbol{\mathcal{B}}=\{\mathcal{B}_1,\mathcal{B}_3\}$ and $\boldsymbol{j}=\left\{0,j_3 \right\}$. The latter limits the kinematics to that of a pure Mode-I fracture. It is possible to see how the dimension reduction of the pairwise potential function generates a hierarchy of terms characterizing the strain energy stored inside the planar element.\\
In \ref{appendix2}, following the microstructural interpretation given in section \ref{ch:motivations} and through the definition of a paradigmatic case of discrete peridynamics, a simple tool for the physical interpretation of the various terms in the reduced energy of our peridynamic continuum is presented. Thanks to the paradigmatic case, it is possible to give an immediate physical interpretation to the terms of (\ref{eq:wred}) that are scaling with the square $p_1$ and the fourth power $p_2$ of the thickness, that is the former are membrane terms and the latter bending. \\
From an analysis of the expression of $p_1$ (see eq. (\ref{eq:p1})), since $\mathcal{A}_1(x_1)$ is the in-plane component of the displacement field for the points on the reduction plane, it can be confirmed that the terms scaling with H$^2$ of $\omega_{\textup{red,a}}$ can be regarded as purely membrane. In the higher-order term, on the contrary, such as $p_2$ (equation \ref{eq:p2}) which multiplies H$^4$, one can assess the presence of purely bending contributions (for example, those depending solely on $\mathcal{B}_1(x_1)^2$), but also of mixed ones. The mixed terms introduce the coupling of membrane behavior and bending behavior. \textit{This is a unique feature of the nonlocal formulation}.
Indeed, coupling between the membrane and bending behaviors is a feature not easily recoverable in local theories, as shown in \ref{appendix3}. In Section \ref{ch:convergence} it is shown that the coupling is lost when the peridynamic model is localized, namely the reduced energy is evaluated in the limit of vanishing horizon $\delta$.

\noindent Lastly, the terms scaling with H$^{6}$ are higher order ones depending only on $\mathcal{B}_3(x_1)$, which is the nonlocal equivalent of a strain deformation through the thickness $\partial _{x_3}(\boldsymbol{u}_a\cdot \boldsymbol{e}_3)$, where $\boldsymbol{e}_3$ is the unit vector normal to the plane $(x_1,x_2)$. We note that in order to recover the kinematics of the Kirchhoff plate theory $\mathcal{B}_3(x_1)$ must be null.

\noindent The contribution of the jump part of the displacement field to the reduced energy, reflected in $\omega_{\textup{red,J}}$, is more scattered. We see contributions of the jump field to both membrane, mixed and bending-related quantities. Here, again, coupling occurs between the different fields of the jump part of the displacement and the continuous part. 
The highest order term in the thickness (H) is determined by the order of the truncation in the Taylor expansion of the continuous part of the displacement. By retaining only terms up to the first order in $x_3$, the highest power becomes 6. This particular choice was made in order to check the convergence of the nonlocal model, which will be done in the last section of this work.

\section{Model implementation and applications} \label{ch:comparison}
Under the aforementioned conditions, the solution for the Euler-Lagrange system of equations (\ref{ELeqs}) of the PD model was achieved by using a Galerkin approach, resulting in a system of the kind
\begin{equation}
\int \limits _{\mathcal{B}_{\alpha}}\left(2\frac{\partial \Lambda_{\mathcal{E}}}{\partial \boldsymbol{q}}-\frac{\partial \Lambda_{\mathcal{W}}}{\partial \boldsymbol{q}}\right)\boldsymbol{\delta q}=0\ ,
\label{eq:reducedlagrangian2}
\end{equation}
which can be solved iteratively.\\
A Mathematica code has then been developed in order to test the model under different loading conditions. We here present displacement-induced tests for symmetric and non-symmetric load distributions. \\
The results have shown that the reduced model is capable of reproducing both traditional and unconventional mechanical behavior such as distal crack nucleation and loss of symmetry in the crack pattern. 
\begin{table}[ht]
\centering
\footnotesize
\begin{tabular}{ |c|c| } 
 \hline
 \textbf{Mechanical and geometric quantities} & \textbf{Value} \\ \hline \hline
 Young's Modulus [MPa] & 5000 \\ 
 Critical surface energy $G_c$ [$\textup{J}/\textup{m}^2$] & 5.3 \\
 Thickness over length (H/L) & 1/25 \\ \hline \hline
 \textbf{Nonlocal parameters} & \textbf{Value} \\ \hline \hline
 Horizon($\delta$)/Length($L$) & 1/5 \\
 Bond constant $c$ [N/mm$^6$] & $7.7\mathrm{x}10^{6}$\\
 Critical stretch $s_{cr}$ [-] & $0.2\mathrm{x}10^{-4}$\\
 \hline
\end{tabular}
\caption{\footnotesize Parameters of the local equivalent material and geometry of the plate (up); nonlocal parameter of the peridynamic bond-based reduced model (down).}
	\label{tab:bent}
\end{table}
\begin{figure}[!ht]
\footnotesize 
\renewcommand{\figurename}{\footnotesize{Figure}}
\centering
	\includegraphics[width=0.49\textwidth]{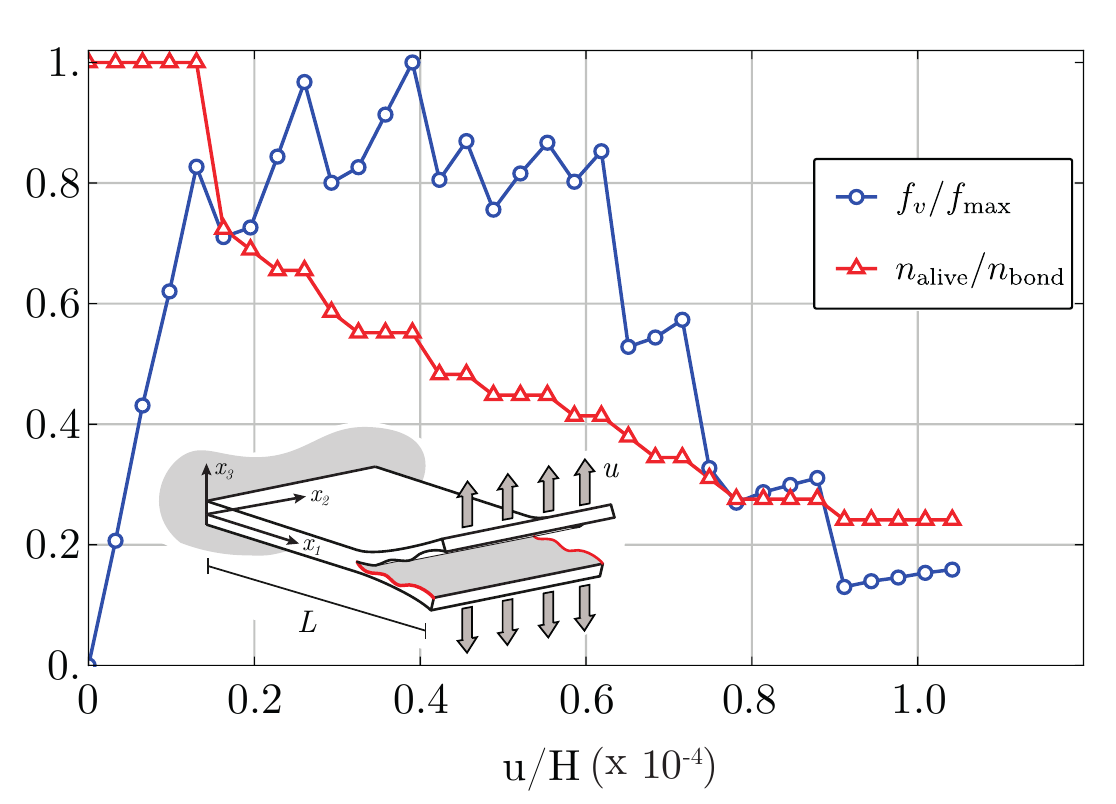}
        \includegraphics[width=0.49\textwidth,angle=0]{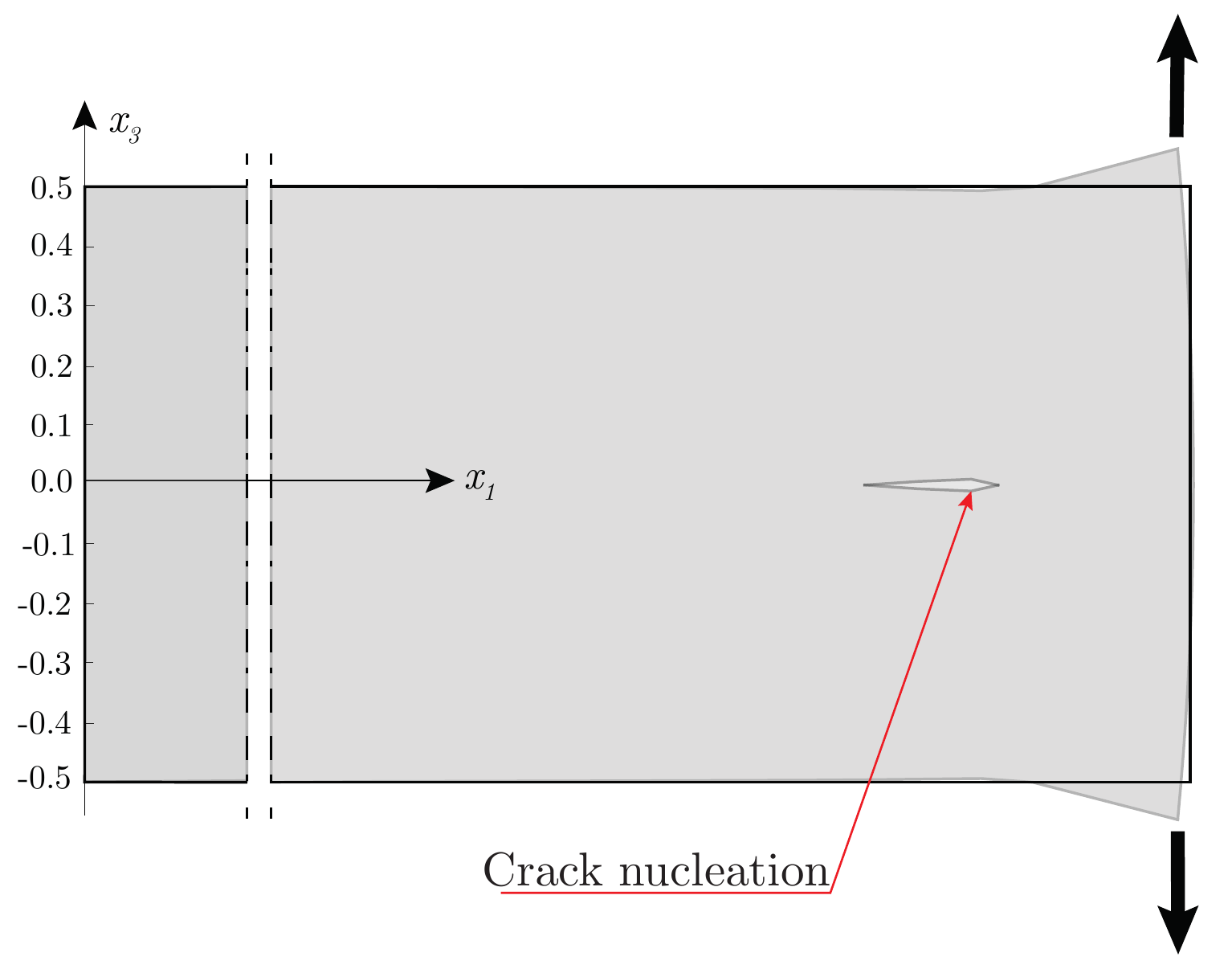}
	\includegraphics[width=0.9\textwidth,angle=0]{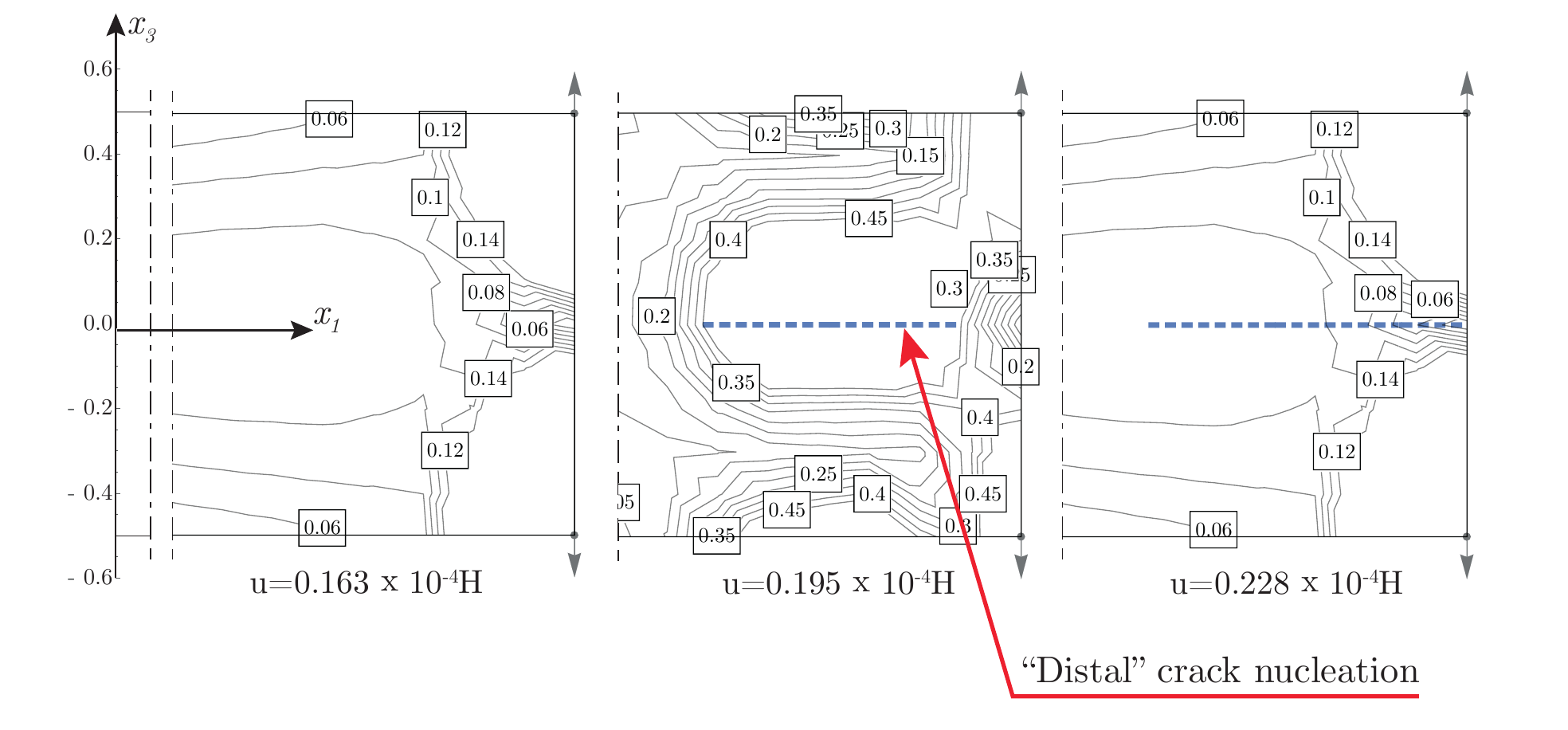}
	\vspace{-5mm}
\caption{\footnotesize Results of the analysis of a peridynamic cantilever plate subjected to two opposing vertical displacements at the upper and lower edges using the present formulation. In the upper part: on the left, the evolution of the force-displacement response and the schematic representation of the test carried out, while on the right, the deformed shape corresponding to an imposed displacement of $u/H=0.195 \times 10^{-4}$ amplified by a factor of 100. In the lower part: level curves representing the displacements - normalized with respect to the imposed one - showing the evolution of crack surface $h(x)$, represented by the dashed blue line, as the imposed displacements at the edges increase.}
\label{anVSnum}
\end{figure}

\subsection{Displacement-induced peeling test} 
\begin{figure}[!ht]
\footnotesize 
\renewcommand{\figurename}{\footnotesize{Figure}}
\centering
	\includegraphics[width=1.4\textwidth,center]{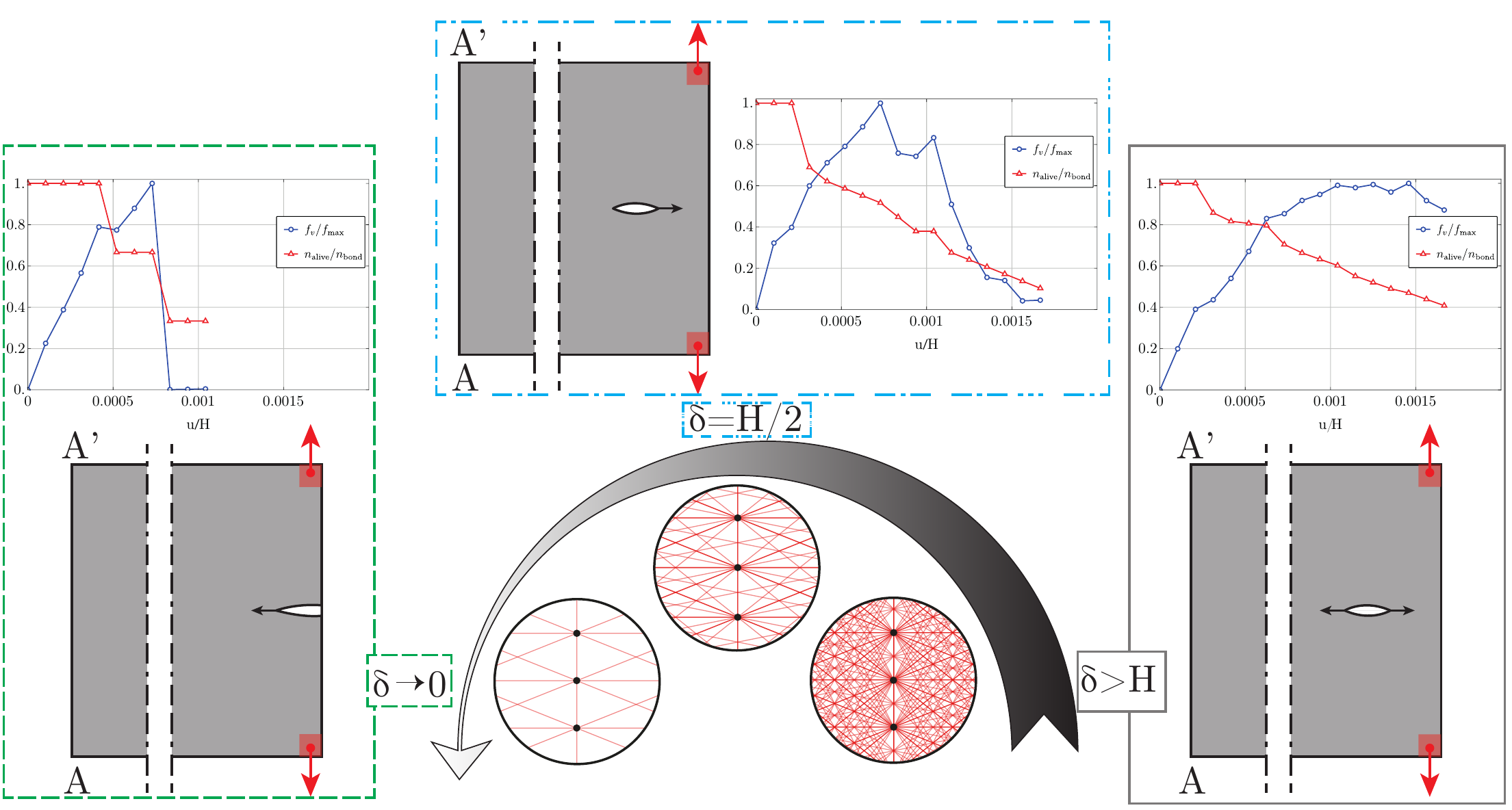}
	\vspace{-4mm}
\caption{\footnotesize Qualitative behavior of crack nucleation and propagation for a peridynamic reduced cantilever plate subjected to a displacement-induced test for a decreasing horizon. The plates are clamped at cross-section AA' and loaded at the opposite edge.}
\label{fig:parametric}
\end{figure}

The peridynamic reduced formulation proposed above is used to study the case of displacement-controlled test inducing through-thickness fracture of a cantilever plate. As shown in Figure \ref{anVSnum}, the plate is loaded by the application of a pair of vertical displacements to the upper and lower part of the free edge. The initial geometry necessitates neither an a priori crack nor a notch in order to develop a crack. This is due to the damage being implemented at the constitutive level in the peridynamic theory and to the kinematic assumptions on the damage-fracture transition taken before.\\
The test has been carried out until a final vertical displacement of around H/600, a quantity which is sufficient for the development of fracture for the chosen elastic and critical parameters (see Table \ref{tab:bent}). In Figure \ref{anVSnum} (above), in blue, the normalized force vs displacement plot is presented, while in red is the fraction of bonds that have yet to break near the loaded area. The latter has been used to investigate the propagation of damage before and during fracture growth. In particular, both damage and fracture surfaces first develop at a \textit{distal section} from the plate edge (loci of the applied load) as shown in Figure \ref{anVSnum} (below), and then propagate in both directions, as observed, e.g., in laminated paper \cite{CONTI2021}.
This unusual response is obtained for a significant nonlocal character of the peridynamic continuum, i.e. horizon larger than the thickness of the plate. In fact, by reducing this parameter the interaction becomes more local and a different response is obtained where the crack nucleates closer to the free edge, ultimately reaching it in the limit of vanishing horizon which is a typical result of standard local continuum theories. Figure \ref{fig:parametric} shows the results of a parametric analysis of $\delta$ ringing from a value of twice the thickness H down to approximately zero, the value at which the fracture is nucleating and propagating from the cross-section at the free edge (where the load is applied).
\noindent When in a peridynamic discrete body or continuum the horizon is reduced, the number of total interactions, i.e. the bonds, of a point is also reduced. As a consequence, the behavior of the structure becomes less cohesive; this feature is clearly shown in the force-displacement plots of the various cases depicted in Figure \ref{fig:parametric}. Surprisingly, the cohesive trait is not completely lost in the local case as is shown in the next Section. Finally, the red lines in the force plots are the relative number of broken bonds, thus they represent the total damage in the zone of the load application.\\
Along with the loss in cohesiveness, a reduction in the number of bonds is due to affect the overall stiffness of the plate. We hence display the result of a comparison of the plate behavior for different horizon sizes, given a constant overall stiffness. This condition can be obtained by increasing the bond constant $c$ of the peridynamic model as $\delta$ decreases; having in mind the paradigmatic micro-structure of a peridynamic discrete body, an increase in $c$ is achieved by thickening each beam that represents a bond, see Figure \ref{fig:comparison} (see \ref{appendix2}).
In the same figure, the comparison shows that the nonlocal micro-structure is capable of absorbing more energy, displaying thus superior toughness when compared with the cases of smaller horizons, which are in this sense more brittle. The relationship between horizon size and total dissipated energy seems to be less than linear as, from our study, an increase of four times the volume of interaction has brought about an increase of total energy dissipated by 1.5 times. 
\begin{figure}[!ht]
\footnotesize 
\renewcommand{\figurename}{\footnotesize{Figure}}
\centering
	\includegraphics[width=\textwidth,center]{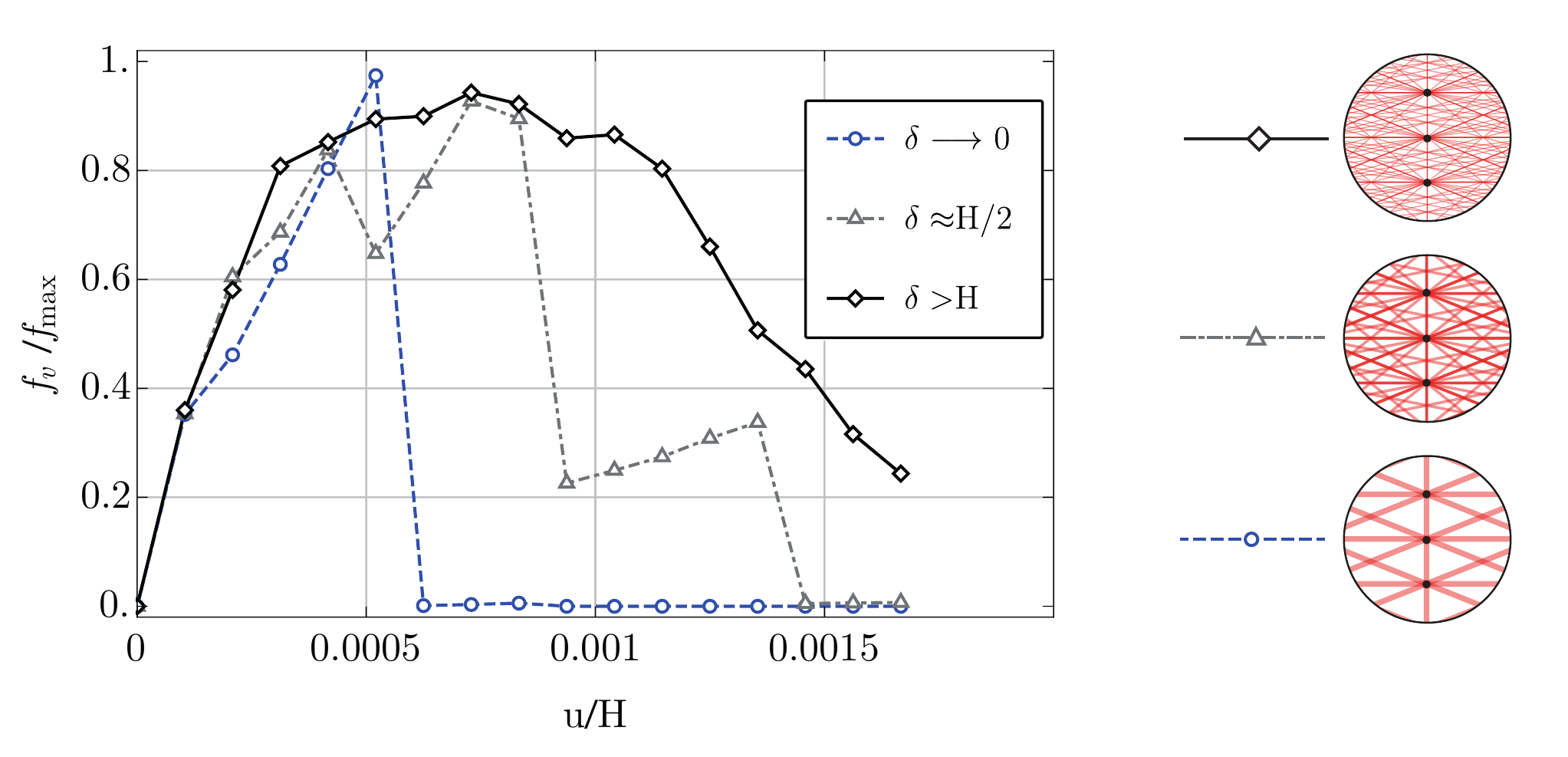}
	\vspace{-4mm}
\caption{\footnotesize Comparison of the force-displacement response of nonlocal plates with different horizon sizes but equal overall stiffness (on the left). On the right, is the equivalent micro-structure for the peridynamic body in the different cases; the increase in bond stiffness is achieved by thickening the cross-section of each beam.}
\label{fig:comparison}
\end{figure}
It is worth mentioning here that contour plots and force displacements plots all depict early-stage crack propagation phenomenon in the peridynamic plate, that is the damaging onset, the nucleation of fracturing embryos and the crack advancement in the very close regions.

\subsection{Non-symmetric load distribution inducing a through-thickness crack}
\label{sec:non-sym}

\begin{figure}[!ht]
\footnotesize 
\renewcommand{\figurename}{\footnotesize{Figure}}
\centering
	\includegraphics[width=\textwidth]{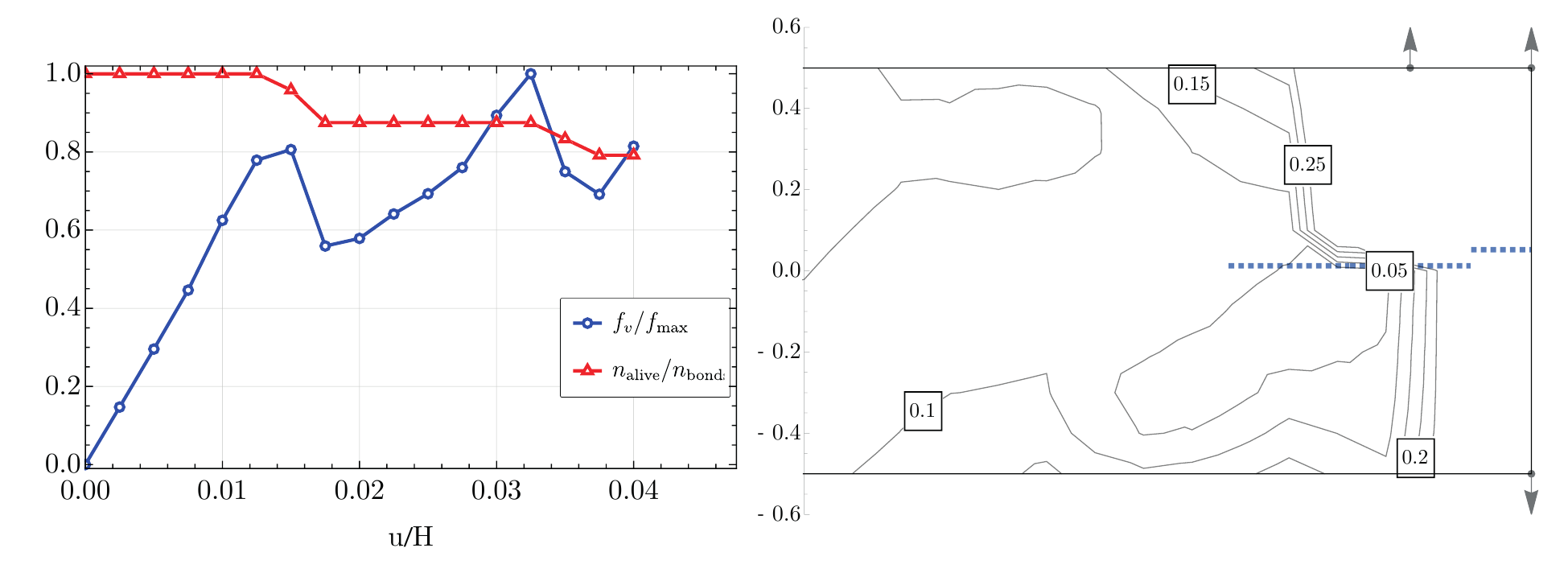}
	\vspace{-4mm}
\caption{\footnotesize Non-symmetrical load distribution leading to a loss of symmetry of the crack path. On the left, the force-displacement plot (in blue) normalized with respect to the peak force, and the number of intact bonds is in red. The latter is representative of damage evolution. On the right, the crack surface (blue dotted line) in the sample loaded by the non-symmetrical distribution of forces and displacements expressed in terms of level set curves.}
\label{fig:ex_2}
\end{figure}

Starting from the previous case of a symmetrically loaded plate, we here explore the effects of an asymmetry in the application of the loads on the crack surface of a cantilever plate. The non-symmetric loading condition is achieved by pulling the upper edge in multiple points while the lower edge of the plate is still pulled from a single one, see Figure \ref{fig:ex_2} on the right. The parameters used for the simulation are $c=7618$N/mm$^6$ (bond constant), $\delta=H/2$ (the horizon), $s_{cr}=0.02036$ (critical elongation of a single bond) and the test has been carried out until a final vertical displacement of approximately H/20.\\
As far as the crack path is concerned, the non-symmetric load distribution induces an unexpected non-symmetric crack trajectory which nucleates and propagates from the external section towards the center of the plate, see Figure \ref{fig:ex_2}. This sensitivity of crack path to even slight loss of symmetry in the prescribed boundary conditions is not typically achieved in thin structures obeying Saint Venant's principle.

\subsection{Mode-II fracture propagation} 
\begin{figure}[!ht]
\footnotesize 
\renewcommand{\figurename}{\footnotesize{Figure}}
\centering
	\includegraphics[width=\textwidth]{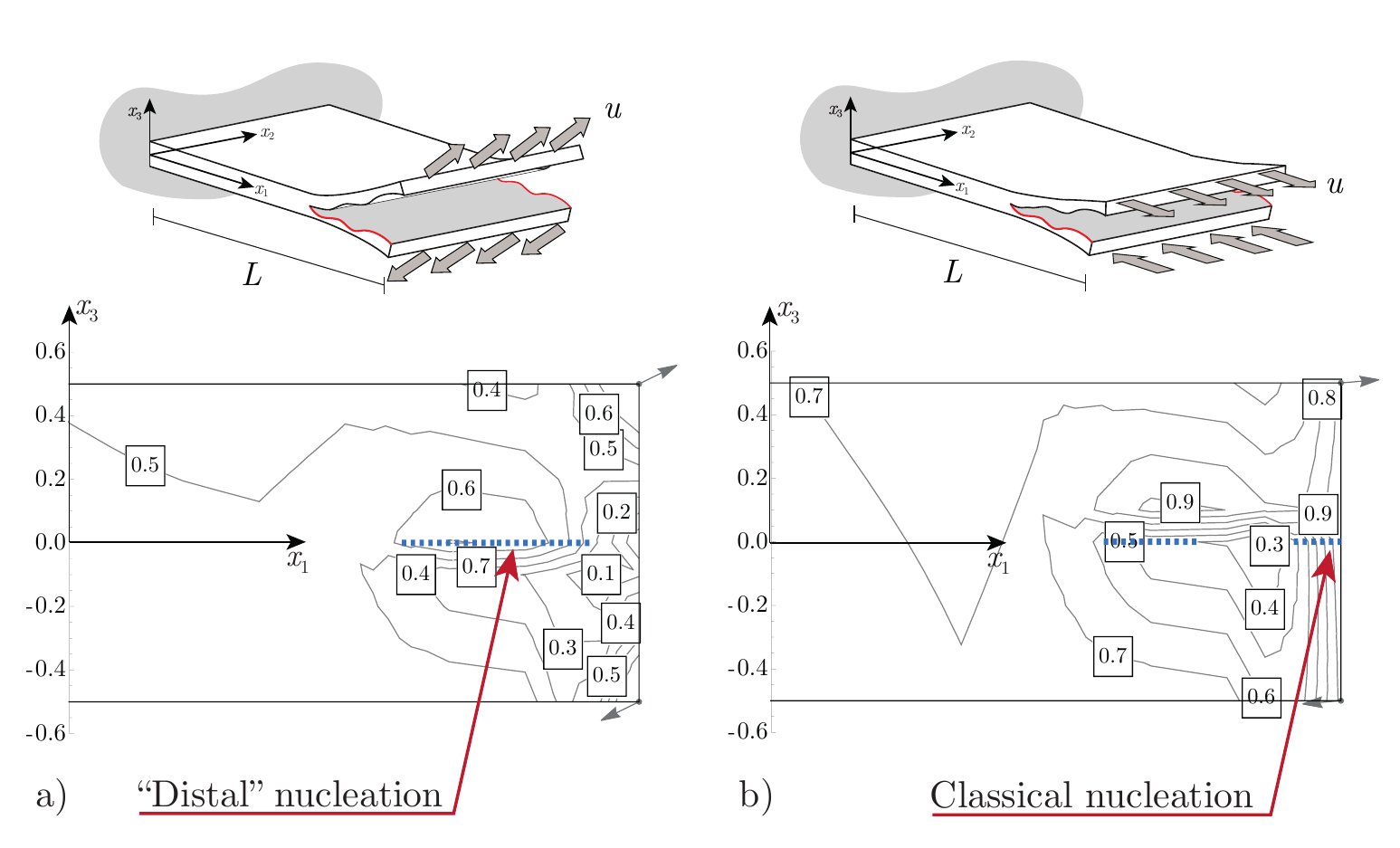}
        \includegraphics[width=0.75\textwidth]{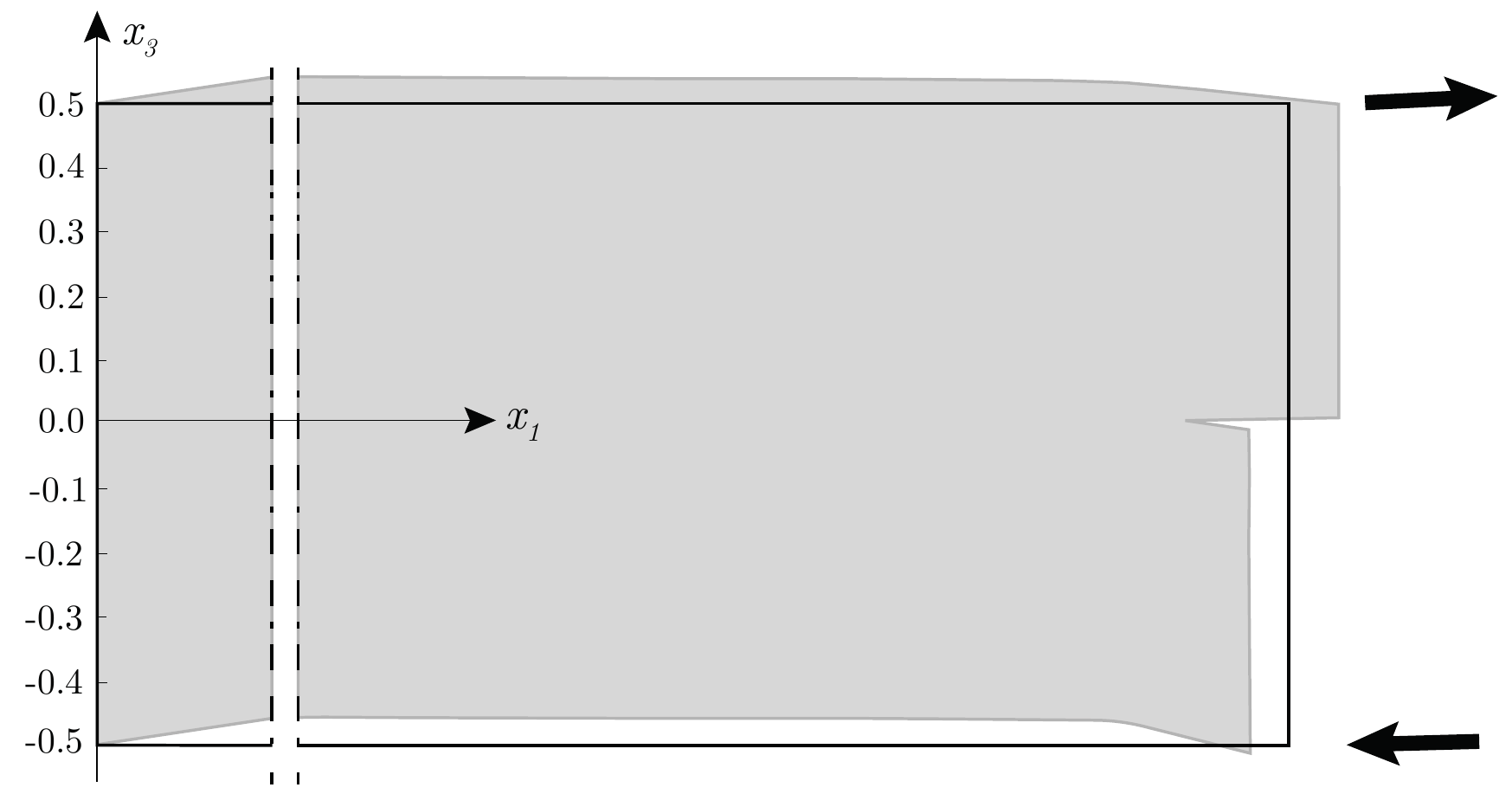}
	\vspace{-4mm}
\caption{\footnotesize A nonlocal peridynamic plate loaded with a couple obtained by two opposing forces of increasing inclination. In the upper part of the figure a schematic representation of the loads and the plate is reported for both the case of a) inclination of the forces of 30° and b) inclination of the forces close to zero. In the central part of the figure are reported the displacements (normalized with respect to the maximum displacement imposed, i.e. $u/H=0.01$) induced in the plate by the forces for the two cases, displaying different crack nucleation and growth (blue dotted line). Lastly, in the lower part of the Figure, it is depicted the deformed shape of the nonlocal plate corresponding to an imposed horizontal displacement of $u/H=0.003$ (with an amplification of 40 times).}
\label{fig:bending}
\end{figure}

Nonlocal peridynamic plates can show loss of continuity through the thickness due to the action of an external couple. To show this, a clamped peridynamic plate is loaded through the application of two opposing forces, applied at the upper and lower edge of the free end of the plate, with a growing inclination (see Figure \ref{fig:bending}). The mechanical and geometrical parameters chosen for the simulation are the same as the previous case shown in Section \ref{sec:non-sym}.
Upon reaching a condition of forces almost horizontal (zero inclination, Figure \ref{fig:bending} on the right) the characteristic distal nucleation shown for the previous case of opposing vertical forces and Mode-I failure is lost and a more \lq \lq classical" crack growth is exhibited with nucleation occurring at the free-end section in a Mode-II fashion. Nonetheless, in the latter case, the evolution of the crack is not continuous and, at a later stage, a more distal crack nucleates far from the first.\\
As a last observation it is useful to highlight that from the various examples presented above it emerges a complex and rich interaction between the applied loads and the displacement field. This makes it very hard to substitute a specific load distribution with a possible static equivalent, such as the resultant, to be applied to a dimensionally reduced plate. The need to follow fracturing/delamination processes makes forces with overall vanishing resultants as relevant as not vanishing ones, the nonzero force and couple resultants being thus not the sole effective loads to be considered. Indeed, the two opposite forces of the first example, applied at the same free surface of the plate along the same vertical direction, give zero global resultant but are however very relevant for delamination, consistently with the classical peeling tests.

\section{Convergence to a local elastic model} \label{ch:convergence}
Convergence of the proposed PD model to local elasticity is assured for the continuous part of the displacement field only \cite{BELLIDO2015,BELLIDO2020,SILLING2008,MENGESHA2015}. Nonetheless, with appropriate scaling of the jump field functions, convergence for vanishing horizon leads to a bounded form of the energy.\\
By applying (\ref{eq:reldisp}) to (\ref{eq:energyPD}), the first term of (\ref{eq:lagrangian2}), which is the elastic energy of a bond-based PD body, becomes
\begin{equation}
\frac{1}{4}\frac{c}{\sigma(\boldsymbol{\xi})} \mu \int \limits _{\mathbb{B}} \int \limits _{\mathbb{H}}\left[(\boldsymbol{\xi} \cdot \boldsymbol{\eta}_{a})^2+2(\boldsymbol{\xi} \cdot \boldsymbol{\eta}_{a})(\boldsymbol{\xi} \cdot \boldsymbol{\eta}_{J})+(\boldsymbol{\xi} \cdot \boldsymbol{\eta}_{J})^2\right] \textup{d}V'
\textup{d}V\ .
\label{expanded}
\end{equation}
To ensure convergence for vanishing nonlocality, i.e. $\delta \rightarrow 0$, the scaling of each term must be checked.

\subsection{Peridynamic parameter evaluation}
Isotropic homogeneous linear elastic materials in the bond-based peridynamic theory are characterized by one single constant, called the bond constant $c$. This is due to the fact that the value of the Poisson's ratio $\nu$ for a bond-based material is fixed to $1/4$ or $1/3$ depending on the dimension of the problem, leaving only one parameter tunable. The constant is typically defined by means of an energetic equivalence with standard local elastic material. It has huge effects on the value of this constant under what conditions this equivalence is imposed, i.e. isotropic expansion, pure elongation or even shear. The energy obtained after the convergence to the local model is, in fact, affected by the choice of the bond constant to the point that certain terms can converge to classical ones typical of local theories while others may not. For example, if one were to make the choice of imposing equivalence of the stretching energy in the peridynamic model and in the local elastic one, only first-order terms in H of the localized PD model would converge while quadratic, cubic and higher-order ones would not.

Commonly, for the evaluation of the bond constant through energy equivalence with local continua, the choice of isotropic expansion is made for the deformation map. This choice is not expected to make all the terms converge to the classical ones, but it can give a general idea of the possibilities of the model obtained by the convergence.
The energy density for a bond-based PD linear elastic material under isotropic expansion ($\boldsymbol{\eta}=\alpha \boldsymbol{\xi}$) is defined as
\begin{equation}
W_{PD}=\frac{1}{2} c\alpha^2\ \int \limits _{\mathbb{H}}|\boldsymbol{\xi}|^{4-b} \textup{dV}=\frac{1}{2} c\alpha^2\ \gamma(\mathbb{H},b,d)\ \delta^{4-b+d}, \label{PDenergyc}
\end{equation}
where $\gamma$ is a scalar function which depends on the shape of the family $\mathbb{H}$, the dimension of the problem d, and the parameter $b$ which comes from the choice of $\sigma(|\boldsymbol{\xi}|)=|\boldsymbol{\xi}|^b$. Likewise, the energy of an isotropic expanding linear elastic material in classic local elasticity is defined as
\begin{equation}
W_{CL}=\frac{1}{2}\alpha^2\ \boldsymbol{I}\cdot \mathbb{E}[\boldsymbol{I}]=\frac{1}{2}\alpha^2\frac{3\textup{E}}{1-2\nu}, \label{energylocalE}
\end{equation}
where $\boldsymbol{I}$ is the identity tensor, while $\mathbb{E}$ is the fourth-order elasticity tensor, E is Young's modulus of the local elastic material and $\nu$ is Poisson's ratio. \\
By enforcing equivalence between the energies (\ref{PDenergyc}) and (\ref{energylocalE}) one recovers 
\begin{equation}
    c=\frac{3 \textup{E}}{\gamma(b)\ \delta^{4-b+d}(1-2\nu)}.
 \label{eq:bondconst}
\end{equation}
In the case of a spherical horizon, one obtains
\begin{equation}
    c=\frac{15\ \textup{E}}{56\ \delta^6}.
    \label{eq:bondconst2}
\end{equation}
According to (\ref{eq:bondconst}), the scaling of the bond constant is then defined as
$c\sim \delta^{b-4-d}$.

\subsection{Displacement scaling}
The continuous part of the energy (the first term of equation (\ref{expanded})) is found to be scaling as
\begin{equation}
\int \limits _{\mathbb{B}} \int \limits _{\mathbb{H}} \frac{c}{|\boldsymbol{\xi}|^b}(\boldsymbol{\xi} \cdot \boldsymbol{\eta}_{a})^2\textup{dVdV}'\sim \delta^{-(4-b+d)-b+2(1+m)+d}=\delta^{2(m-1)},
\label{continuousdelta}
\end{equation}
since the scaling of $c$ is defined in (\ref{eq:bondconst2}), and the other terms scale as follow: $|\boldsymbol{\xi}| \sim \delta$; $|\boldsymbol{\eta}| \sim \delta^m$; $V' \sim \delta^d$. For the integral term to stay bounded and nonvanishing one requires $m=1$.\\
Hence, $\boldsymbol{\eta}_{a}\sim \delta^1$, which means that
\begin{equation}
\boldsymbol{\eta}_{a}\approx \nabla_{\boldsymbol{x}} \boldsymbol{\mathcal{A}}(x_1,x_2)\ \boldsymbol{\xi}+\nabla_{\boldsymbol{x}} \boldsymbol{\mathcal{B}}(x_1,x_2) \boldsymbol{\xi}\ x_3+\boldsymbol{\mathcal{B}}(x_1,x_2)\ \boldsymbol{\xi}\cdot \boldsymbol{e}_3 + \cdots
\end{equation}
defines the scaling of the shape functions, since $\boldsymbol{\xi}\sim \delta^1$. In particular, no scaling is required
\begin{equation}
\nabla_{\boldsymbol{x}} \boldsymbol{\mathcal{A}}(x_1,x_2) \sim \delta^0\ \ ,\ \ \nabla_{\boldsymbol{x}} \boldsymbol{\mathcal{B}}(x_1,x_2)\sim \delta^0\ \ ,\ \ \boldsymbol{\mathcal{B}}(x_1,x_2)\sim \delta^0.
\end{equation}
In a similar fashion, the second term of (\ref{expanded}) must follow the following scaling: 
\begin{equation}
\int \limits _{\mathbb{B}} \int \limits _{\mathbb{H}} \frac{c}{|\boldsymbol{\xi}|^b}(\boldsymbol{\xi} \cdot \boldsymbol{\eta}_{J})(\boldsymbol{\xi} \cdot \boldsymbol{\eta}_{a})\textup{dVdV}'\sim \delta^{-(4-b+d)-b+(2+m+n)+d}=\delta^{-2+m+n}. 
\end{equation}
In order for the energy to stay bounded the scaling of the jump part of the displacement field (defined by $n$) must fulfill the condition of $n\geq 1$, since from (\ref{continuousdelta}) $m=1$. Accordingly, from the last term of the energy one obtains:
\begin{equation}
\int \limits _{\mathbb{B}} \int \limits _{\mathbb{H}} \frac{c}{|\boldsymbol{\xi}|^b}(\boldsymbol{\xi} \cdot \boldsymbol{\eta}_{J})^2\textup{dVdV}'\sim \delta^{-(4-b+d)-b+2(1+n)+d} =\delta^{2(n-1)},
\label{rescalejump}
\end{equation}
which gives the redundant condition: $n\geq 1$. 
In view of (\ref{eq:jumpdispl}) and for vanishing nonlocality one can approximate the relative jump displacement $\boldsymbol{\eta}_J$ as
\begin{equation}
\boldsymbol{\eta}_{J}\approx \left.\nabla_{\boldsymbol{\xi}} \boldsymbol{u}'_{J}\right|_{\boldsymbol{\xi}=0}\cdot \boldsymbol{\xi},
\label{etaJlin}
\end{equation}
where $\boldsymbol{u}'_{J}$ is the displacement of the particle $\boldsymbol{x}'$. If we call $p$ the scaling of $\nabla_{\boldsymbol{\xi}} \boldsymbol{u}'_{J}$ then by virtue of (\ref{rescalejump}), $p\geq 0$. Though, since
\begin{equation}
\nabla_{\boldsymbol{x}}\left. \boldsymbol{u}'_{J}\right|_{\boldsymbol{\xi}=0}= \nabla_{\boldsymbol{x}} \boldsymbol{j}\ \Theta(x_3-h(x_1,x_2))+\boldsymbol{j}\otimes\left(\boldsymbol{e}_3-\nabla_{\boldsymbol{x}} h(x_1,x_2)\right)\phi(x_3-h(x_1,x_2)),
\label{etaJlin2}
\end{equation}
where $\phi$ is the Dirac delta distribution, one can easily assess that in order for $h(x_1,x_2)$ and the energy to be bounded, the following scaling must hold
\begin{equation}
\nabla_{\boldsymbol{x}} \boldsymbol{j} \sim \delta^0\ \ , \ \ \boldsymbol{j} \sim \delta^{0}\ \ , \ \  \nabla_{\boldsymbol{x}} h \sim \delta^{0}.
\end{equation}

\subsection{The scaling of the failure criterion}

Alongside the energy, also the damage criterion ($s<s_{\textup{cr}}$) scales as $\delta \rightarrow 0$. The scaling of the critical stretch $s_{\textup{cr}}$ is defined by equation (\ref{griffith}), so $s_{\textup{cr}}\sim \delta^{-1/2}$. The scaling of the stretch $s$, on the other hand,
can be obtained by employing equations (\ref{etaJlin}) and (\ref{etaJlin2})
\begin{equation}
s=\frac{\boldsymbol{\xi}\cdot \left.\nabla_{\boldsymbol{\xi}} \boldsymbol{u}'_{J}\right|_{\boldsymbol{\xi}=0}\cdot \boldsymbol{\xi}}{|\boldsymbol{\xi}|^2}=\left.\nabla_{\boldsymbol{\xi}} \boldsymbol{u}'_{J}\right|_{\boldsymbol{\xi}=0}:\frac{\boldsymbol{\xi}\otimes \boldsymbol{\xi}}{|\boldsymbol{\xi}|^2}.
\end{equation}
The second tensor in the double dot product\footnote{Given two second-order tensors, A and B, we mean by double dot product the operation A$:$B$^T=\textup{Tr}(\textup{AB})=A_{ij}B_{ji}$.} is a quantity that scales as $\delta^0$ whereas the first tensor harbors a singularity, the Dirac's Delta function $\phi$, which for $x_3=h(x_1,x_2)$ makes the stretch infinite. Hence, whenever on the crack surface, the criterion is immediately not satisfied. Finally:
\begin{equation}
s<s_{\textup{cr}}\rightarrow \left\{
\begin{tabular}{l l}
$\textup{False}$&\textup{for} $x_3=h(x_1,x_2)$\\
True &\textup{otherwise}
\end{tabular}
 \right.
 \label{failurelocal}
\end{equation}

\subsection{Localized energy in plane strain}

localization of the PD non-local model has been obtained by means of a limit operation, for vanishing $\delta$, on the PD non-local elastic energy. 
The localized energy obtained in this way is composed of a part entirely defined by the continuous part of the displacement field, the term (\ref{continuousdelta}), and a part composed by mix and purely jump terms
\begin{equation}
\mathcal{E}_{local}=\mathcal{E}_{local,a}+\mathcal{E}_{local,J}, \label{localtotal}
\end{equation}
where for the assumption of continuous displacement field (\ref{continuousfield}) truncated at first order in $x_3$, and plane strain
\begin{equation}
    \begin{split}
    \mathcal{E}_{local,a}=& \textup{HE}\left(\frac{3}{56} \mathcal{A}_1'(x_1)^2+\frac{5}{84}
   \mathcal{B}_3(x_1) \mathcal{A}_1'(x_1)+\frac{5}{168}
   \mathcal{A}_3'(x_1)^2+\right.\\
   &\ \ \ \ \ \ \left. \frac{5}{84} \mathcal{B}_1(x_1) \mathcal{A}_3'(x_1)+\frac{5}{168}
   \mathcal{B}_1(x_1)^2+\frac{3}{56} \mathcal{B}_3(x_1)^2\right)+\\
   &\textup{H}^3 \textup{E}\left(\frac{1}{224} \mathcal{B}_1'(x_1)^2+\frac{5
   \mathcal{B}_3'(x_1)^2}{2016}\right).
   \end{split}
   \label{eq:localenA}
\end{equation}
\noindent Here $\boldsymbol{\mathcal{A}}=\{\mathcal{A}_1,\mathcal{A}_3\}$, $\boldsymbol{\mathcal{B}}=\{\mathcal{B}_1,\mathcal{B}_3\}$ and the primes indicates derivative with respect to $x_1$. Meanwhile,
\begin{align} \label{eq:localenJ}
    \mathcal{E}_{local,J}=&\textup{H}\textup{E}\ \left(\frac{5}{168} \mathcal{B}_1(x_1)j_3'(x_1)+\frac{5}{336}j_3'(x_1){}^2+\frac{5}{168}j_3'(x_1)\mathcal{A}_3'(x_1)\right)+  \\
    & \textup{H}^2\text{E} \left(\frac{5}{672} j_3'(x_1) \mathcal{B}_3'(x_1)\right) + \nonumber\\
    & \text{E} \ \ \left(\frac{3}{28} j_3(x_1) \mathcal{B}_3(x_1)-\frac{5}{84} j_3'(x_1) \mathcal{B}_1(x_1)\mathit{h}(x_1)-\frac{5}{168} \mathit{h}(x_1) j_3'(x_1){}^2+\right.\nonumber\\
        &\ \ \ \ \left. -\frac{5}{84} \mathit{h}'(x_1) j_3(x_1)\mathcal{B}_1(x_1)-\frac{5}{84} j_3(x_1) j_3'(x_1) \mathit{h}'(x_1)\right. +\nonumber\\
        &\ \ \ \ \ -\frac{5}{84} j_3'(x_1) \mathit{h}(x_1) \mathcal{A}_3'(x_1)-\frac{5}{84} j_3(x_1) \mathit{h}'(x_1) \mathcal{A}_3'(x_1)+\nonumber\\
        &\ \ \ \ \ \frac{5}{84} j_3(x_1) \mathcal{A}_1'(x_1)+\frac{5}{84}\mathit{h}(x_1) j_3(x_1){} \mathcal{B}_1'(x_1)+\nonumber\\
        &\ \ \ \ \ \left. -\frac{5}{168} j_3'(x_1) \mathit{h}(x_1){}^2 \mathcal{B}_3'(x_1)-\frac{5}{84} j_3(x_1) \mathit{h}(x_1) \mathit{h}'(x_1) \mathcal{B}_3'(x_1) \right) \nonumber\ \ ,
\end{align}

\noindent where the hypothesis of purely Mode-I crack development $\boldsymbol{j}=\{0,j_3\}$ has been considered.\\
Equation (\ref{localtotal}) basically represents a material with cohesive constitutive law due to the energy associated with a jump in the displacement.\\
The localized formulation of the nonlocal model introduced in section \ref{ch:pdmodel} can now be obtained by writing the Lagrangian for a local plate where the internal energy is that of equations (\ref{eq:localenA}-\ref{eq:localenJ}) and then minimizing it. 

\subsection{Kirchhoff-like plate under Mode-I fracture} \label{kirchhoffkin}
The kinematics of a Kirchhoff plate is readily recovered by imposing 
\begin{equation}
\mathcal{A}_1(x_1)\rightarrow 0,\ \ \mathcal{B}_{3}(x_1)\rightarrow 0,\ \ \mathcal{B}_1(x_1)\rightarrow -\mathcal{A}_3'(x_1),
\end{equation}
such that
\begin{equation}
\boldsymbol{u}_{a}(x_1,x_3)= \{-\mathcal{A}_3'(x_1)x_3,\ \mathcal{A}_3(x_1) \}.
\end{equation}
Mode I delamination is achieved by choosing the jump function and its derivative in the following way:
\begin{equation}
j_1(x_1)\rightarrow 0, \ \ j'_1(x_1)\rightarrow 0,
\end{equation}
such that
\begin{equation}
\boldsymbol{u}_{J}[x_1,x_3]=\{0,j_3(x_1)\ \Theta(x_3-h(x_1))\},
\end{equation}
where $\Theta$ is the Heaviside function. Under these conditions the terms of localized energy become
\begin{equation}
\mathcal{E}_{local,a}=\frac{1}{224} \textup{E} \textup{H}^3 \mathcal{A}_3''(x_1){}^2, 
\end{equation}
and
\begin{align}
    \mathcal{E}_{local,J}=& \text{E}\ \left(\frac{5}{84}j_3(x_1) j_3'(x_1) \mathit{h}'(x_1)+\frac{5}{168}
   \mathit{h}(x_1) j_3'(x_1){}^2+\right. \\
   &\ \ \ \ \ \ \ \ \ \left. -\frac{5}{84}j_3(x_1) \mathit{h}(x_1) \mathcal{A}_3''(x_1) -\frac{5}{336}j_3'(x_1){}^2\right)\ , \nonumber
\end{align}
respectively. It is worth mentioning that the limiting local energy for the continuous part of the displacement is a quantity resembling the classical result for Kirchhoff plates: $\mathcal{E}_{local,a}\ = \frac{\textup{H}^3 \textup{E}}{12(1-\nu^2)} \mathcal{A}_3''(x_1){}^2$, where for obvious reasons a Poisson ratio of $1/4$ has to be considered.

\section{Conclusions}
In the present paper, a reduced model for the explicit study of though-thickness fracture nucleation and propagation in thin structures is put forward. \\ 
The model is obtained by making a hypothesis on the kinematics of the thin element, which is assumed as the sum of a continuous part and a jump part. A particular choice of these fields is made which expresses the dependence on the out-of-plane variable explicitly, thus making the integration through the thickness feasible. The resulting reduced model retains information on the loss of continuity of the material through the functions defining the jump. The proposed model has distinguished weak planes/surfaces, yet it leads to the recovery of both horizontal and deviated crack patterns across the thickness of the plate.\\
The dimension reduction procedure generates a hierarchy of terms in the elastic energy stored inside the plate. A mechanical interpretation of those terms is possible (especially for the part of the energy associated with the continuous part of the displacement) and is proposed through the definition of a simple paradigm of peridynamic structure. It is found that the hierarchical form of the energy shows the coupling of membrane and bending behavior despite the formulation being expressed in a linear setting. \\
The reduced model is then tested in a symmetric displacement-induced delamination test for a cantilever plate and a broad range of qualitative responses are obtained for a varying horizon. In particular, for a horizon larger than the height of the plate \textit{ distal} nucleation is observed, whereas in the limit of vanishing horizon a classic result of linear fracture mechanics is recovered with the propagation of the fracture starting from the loaded cross-section of the plate. Apart from the crack path development, different horizons have proven to greatly influence the force-displacement response of the structure, leading to superior toughness and energy dissipation in the nonlocal model with a greater horizon and a more brittle behavior in the case of a smaller horizon.\\
A non-symmetrical displacement-induced test is also performed and a relevant sensitivity of the peridynamic plate emerges from the simulation where a curved crack path characterizes the response at failure of the thin nonlocal element. \\
To further investigate the local limit of the model, localization of the nonlocal reduced formulation is performed. Firstly, the convergence of the nonlocal energy to a finite and non-vanishing local equivalent is assessed. The localized reduced model shows a cohesive nature, which is expressed by the fact that energy can be stored by the part of the energy associated with the discontinuous displacement field when a fracture is propagating. 

\section*{CRediT authorship contribution statement}
\textbf{R. Cavuoto}: Developed the theory, performed the calculations and computations, wrote and edited the manuscript. 
\textbf{A. Cutolo}: Performed the computations, wrote and edited the manuscript. 
\textbf{K. Dayal}: Developed the theory, wrote and edited the manuscript, supervised the whole work.
\textbf{M. Fraldi}: Developed the theory, wrote and edited the manuscript, supervised the whole work.
\textbf{L. Deseri}: Developed the theory, wrote and edited the manuscript, supervised the whole work.

\section*{Declaration of Competing Interest}
The authors declare that they have no known competing financial interests or personal relationships that could have appeared to
influence the work reported in this paper.

\section*{Acknowledgements}
LD, AC and MF gratefully acknowledge the support of the Italian Ministry of Research (MIUR) through the grants PRIN-20177TTP3S and PON “Stream”-ARS01\_01182. LD also gratefully thanks the support of the European Commission through (i) FET Open “Boheme” grant no. 863179, and (ii) LIFE GREEN VULCAN LIFE19 ENV/IT/000213, and (iii) ERC-ADG-2021-101052956-BEYOND. 
Kaushik Dayal thanks Army Research Office (MURI W911NF-19-1-0245), Office of Naval Research (N00014-18-1-2528), and National Science Foundation (DMREF 2118945, DMS 2108784) for financial support.

\appendix
\section{Expanded form of the reduced energy} \label{ch:appendix1}
\noindent 
The hierarchical form of the reduced energy is reported here again for the reader:
\begin{equation}
    \omega_{\textup{red}}=\omega_{\textup{red},a}+\omega_{\textup{red},J}
\end{equation}
where in the case of $\phi=1$, specializes to
\begin{align*}
    \omega_{\textup{red},a}&=\textup{H}^2\ p_1(\phi,\boldsymbol{u}_a)+\textup{H}^4\ p_2(\phi,\boldsymbol{u}_a)+\textup{H}^6\ p_3(\phi,\boldsymbol{u}_a)\ \ ;\\
    \omega_{\textup{red},J}&=r_0(\boldsymbol{u}_J)+\textup{H}\ r_1(\boldsymbol{u}_a,\boldsymbol{u}_J)+\textup{H}^2\ r_2(\boldsymbol{u}_J)+\\
    &\ \ \ \ \textup{H}^3\ r_3(\boldsymbol{u}_a,\boldsymbol{u}_J)+\textup{H}^4\ r_4(\boldsymbol{u}_a,\boldsymbol{u}_J)+\textup{H}^5\ r_5(\boldsymbol{u}_a,\boldsymbol{u}_J) \nonumber\ \ \ ,
\end{align*}
where the $p_i$ functions are those specified by \cref{eq:p1,eq:p2,eq:p3}.
On the other hand, the coefficients of the reduced energy associated with the jump part of the displacements are: 
\begin{align*}
    r_0&=-\frac{1}{6} j_3(x_1') j_3(x_1) \mathit{h}(x_1') \mathit{h}(x_1) \left(2 \mathit{h}(x_1')^2-3 \mathit{h}(x_1') \mathit{h}(x_1)+2 \mathit{h}(x_1)^2\right)\ ;\\
    r_1&= -\frac{1}{3} j_3(x_1')       \mathcal{A}_3(x_1') \mathit{h}(x_1')^3-\frac{1}{6} j_3(x_1')^2 \mathit{h}(x_1')^3+\frac{1}{3} j_3(x_1') \mathit{h}(x_1')^3 \mathcal{A}_3(x_1)+\\
        &\ \ \ \ \frac{1}{2} j_3(x_1') \mathit{h}(x_1')^2 (x_1'-x) \mathcal{A}_1(x_1)+\frac{1}{2} j_3(x_1) \mathcal{A}_1(x_1') (x-x_1') \mathit{h}(x_1)^2+\\
        &\ \ \ \ \frac{1}{3} j_3(x_1) \mathcal{A}_3(x_1') \mathit{h}(x_1)^3+\frac{1}{6} j_3(x_1') j_3(x_1) \mathit{h}(x_1')^3+\frac{1}{6} j_3(x_1') j_3(x_1) \mathit{h}(x_1)^3+\\
        &\ \ \ \ \frac{1}{3} j_3(x_1) (x_1'-x) \mathit{h}(x_1)^3 \mathcal{B}_1(x_1)-\frac{1}{4} j_3(x_1') \mathit{h}(x_1')^4 \mathcal{B}_3(x_1')+\\
        &\ \ \ \ \left. -\frac{1}{4} j_3(x_1) \mathit{h}(x_1)^4 \mathcal{B}_3(x_1) +\frac{1}{2} j_3(x_1')
        \mathcal{A}_1(x_1') \mathit{h}(x_1')^2 (x-x_1')\right.+\\
        &\ \ \ \ \frac{1}{2} j_3(x_1) (x_1'-x) \mathcal{A}_1(x_1) \mathit{h}(x_1)^2+\frac{1}{3}  j_3(x_1') \mathit{h}(x_1')^3 (x-x_1') \mathcal{B}_1(x_1')+\\
        &\ \ \ \ -\frac{1}{6} j_3(x_1)^2-\frac{1}{3}  j_3(x_1) \mathcal{A}_3(x_1) \mathit{h}(x_1)^3  \mathit{h}(x_1)^3\ ;\\
    r_2&= -\frac{1}{8} j_3(x_1') j_3(x_1) \left(\mathit{h}(x_1')^2+\mathit{h}(x_1)^2\right)\ ;\\
    r_3&=-\frac{1}{12} j_3(x_1') \mathcal{A}_3(x_1') \mathit{h}(x_1')-\frac{1}{24} j_3(x_1')^2 \mathit{h}(x_1')+\frac{1}{12} j_3(x_1') \mathit{h}(x_1') \mathcal{A}_3(x_1)+\\
        &\ \ \ \ -\frac{1}{12} j_3(x_1) \mathcal{A}_3(x_1) \mathit{h}(x_1)+\frac{1}{8} j_3(x_1') \mathcal{A}_1(x_1') (x_1'-x)+\\
        &\ \ \ \ \frac{1}{8} j_3(x_1) (x-x_1')\mathcal{A}_1(x_1)+\frac{1}{12} j_3(x_1) (x_1'-x) \mathcal{B}_1(x_1') \mathit{h}(x_1)+\\
        &\ \ \ \ \frac{1}{12} j_3(x_1') \mathit{h}(x_1') (x-x_1') \mathcal{B}_1(x_1)-\frac{1}{12} j_3(x_1') \mathit{h}(x_1')^2 \mathcal{B}_3(x_1)+\\
        &\ \ \ \ -\frac{1}{24} j_3(x_1') \mathit{h}(x_1')^2 \mathcal{B}_3(x_1')-\frac{1}{24} j_3(x_1)^2 \mathit{h}(x_1)-\frac{1}{24} j_3(x_1) \mathit{h}(x_1)^2 \mathcal{B}_3(x_1)+\\
        &\ \ \ \ \frac{1}{8} j_3(x_1) \mathcal{A}_1(x_1') (x_1'-x)+\frac{1}{12} j_3(x_1) \mathcal{A}_3(x_1') \mathit{h}(x_1)-\frac{1}{12} j_3(x_1) \mathcal{B}_3(x_1') \mathit{h}(x_1)^2+\\
        &\ \ \ \ \frac{1}{8} j_3(x_1') (x-x_1') \mathcal{A}_1(x_1)+\frac{1}{24} j_3(x_1') j_3(x_1) \mathit{h}(x_1)+\frac{1}{24} j_3(x_1') j_3(x_1) \mathit{h}(x_1')\ ;\\
    r_4&=\frac{1}{12} j_3(x_1') \mathcal{A}_3(x_1')-\frac{1}{12} j_3(x_1') \mathcal{A}_3(x_1)-\frac{1}{12} j_3(x_1) \mathcal{A}_3(x_1')+\\
        &\ \ \ \ \frac{1}{24} j_3(x_1')(x_1'-x) \mathcal{B}_1(x_1) +\frac{1}{24} j_3(x_1) (x-x_1') \mathcal{B}_1(x_1')+\\
        &\ \ \ \ \frac{1}{12} j_3(x_1) \mathcal{A}_3(x_1)+\frac{1}{24}j_3(x_1)^2+\frac{1}{24} j_3(x_1') (x_1'-x) \mathcal{B}_1(x_1')+\\
        &\ \ \ \ -\frac{1}{96} j_3(x_1') j_3(x_1)+\frac{j_3(x_1')^2}{24}+\frac{1}{24} j_3(x_1) (x-x_1') \mathcal{B}_1(x_1)\ ; \\
    r_5&=\frac{1}{48} j_3(x_1) \mathcal{B}_3(x_1')+\frac{1}{48} j_3(x_1') \mathcal{B}_3(x_1)+\frac{5}{192} j_3(x_1') \mathcal{B}_3(x_1')+\frac{5}{192} j_3(x_1) \mathcal{B}_3(x_1).
\end{align*}

\section{A micro-structural interpretation for bond-based PD} \label{appendix2} 
\begin{figure}[ht]
\footnotesize
\renewcommand{\figurename}{\footnotesize{Figure}}
	\begin{center}
	\includegraphics[trim=1 1 1 9,clip,width=0.7\textwidth]{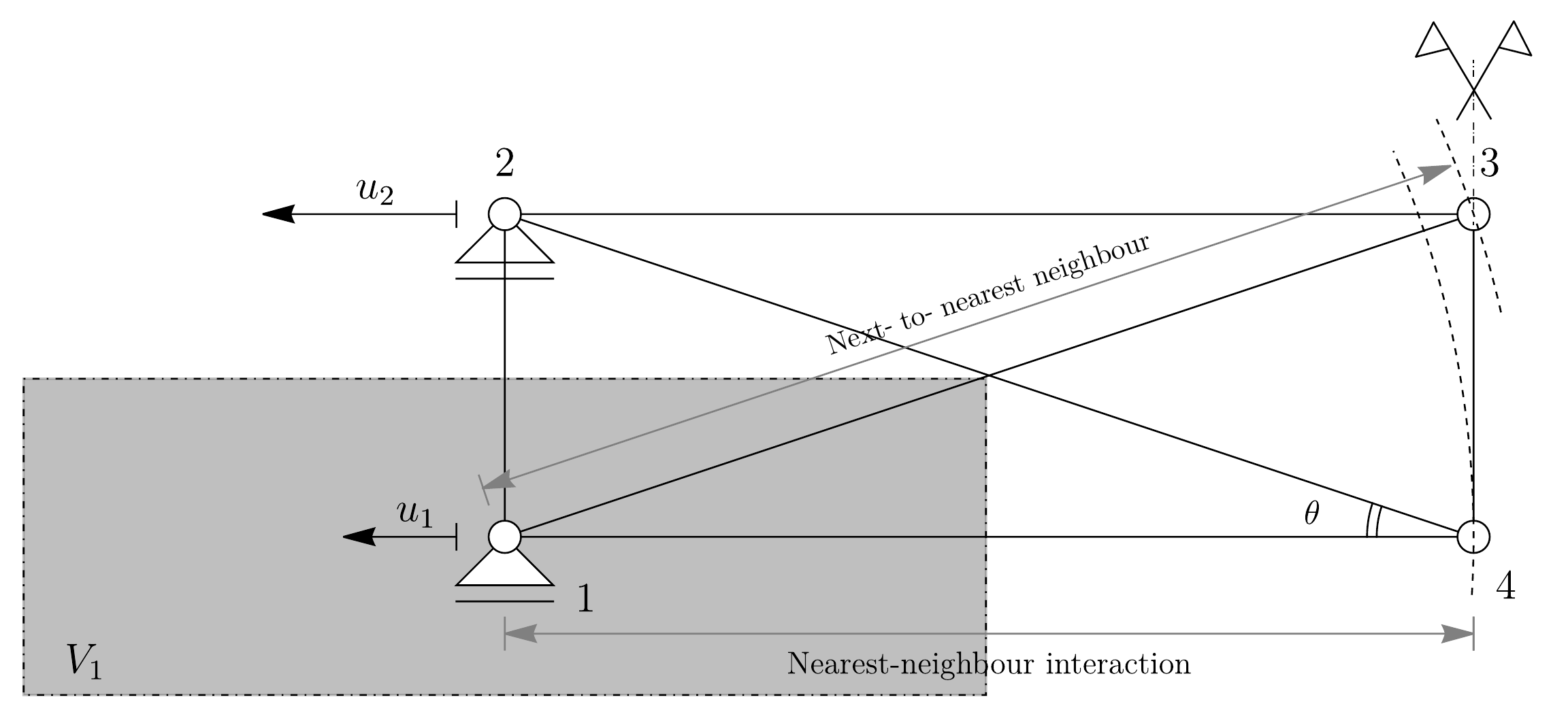}
	\end{center}
	\vspace{-7mm}
\caption{\footnotesize Simplest structure representing a bond-based PD body. The nodes are the material particles of the body, while the bonds are represented by the truss connecting each couple of nodes. The structure has been made symmetric, both in loading conditions that in elements stiffness and coordinates. The gray region (denoted by $V_1$) represents the volume ascribable to node 1. The horizon $\delta$ is taken to be equal to the diagonals of the structure.
}
\label{paradigm}
\end{figure}

\begin{figure}[ht]
\footnotesize
\renewcommand{\figurename}{\footnotesize{Figure}}
	\begin{center}
	\includegraphics[width=0.49\textwidth]{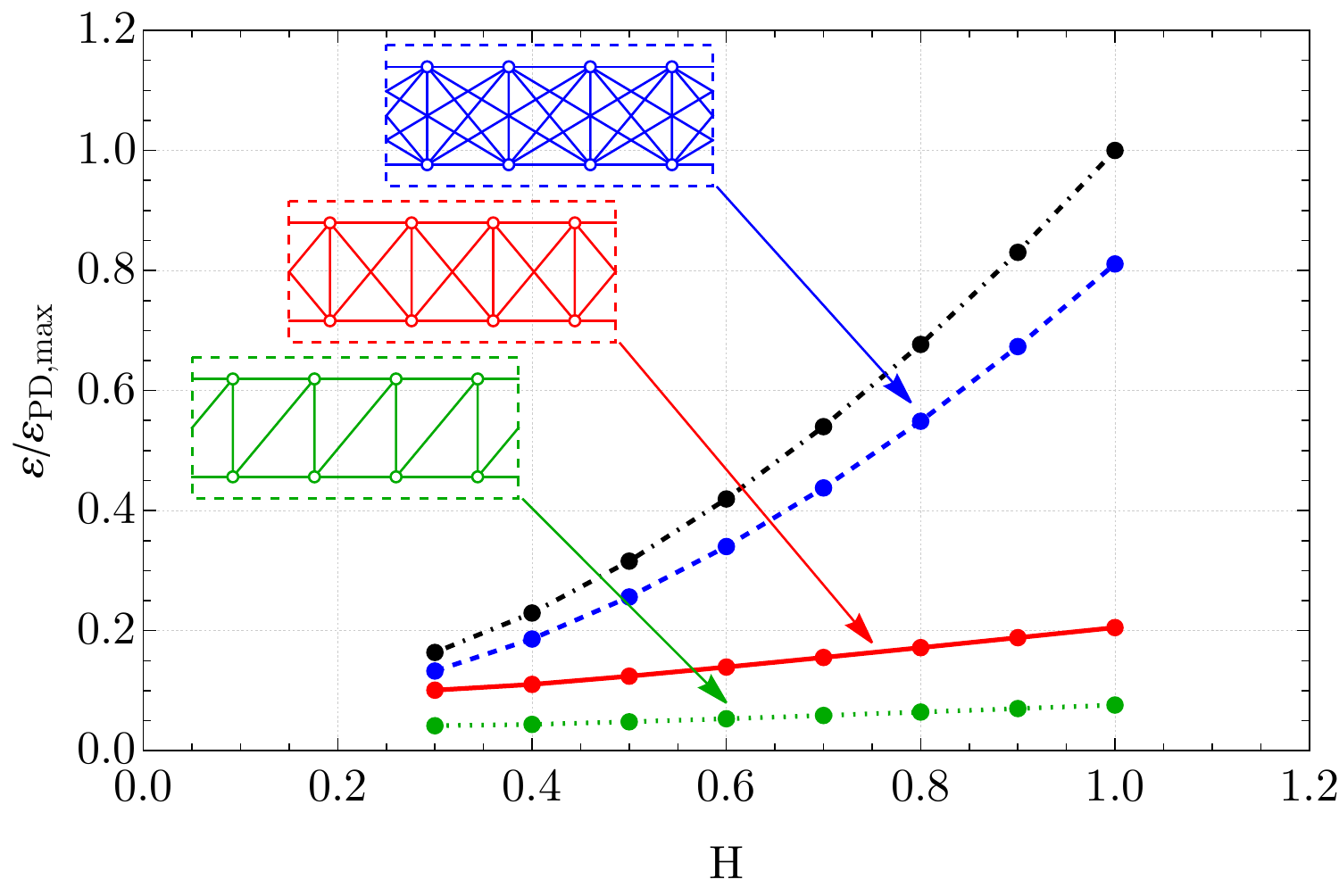}
	\includegraphics[width=0.49\textwidth]{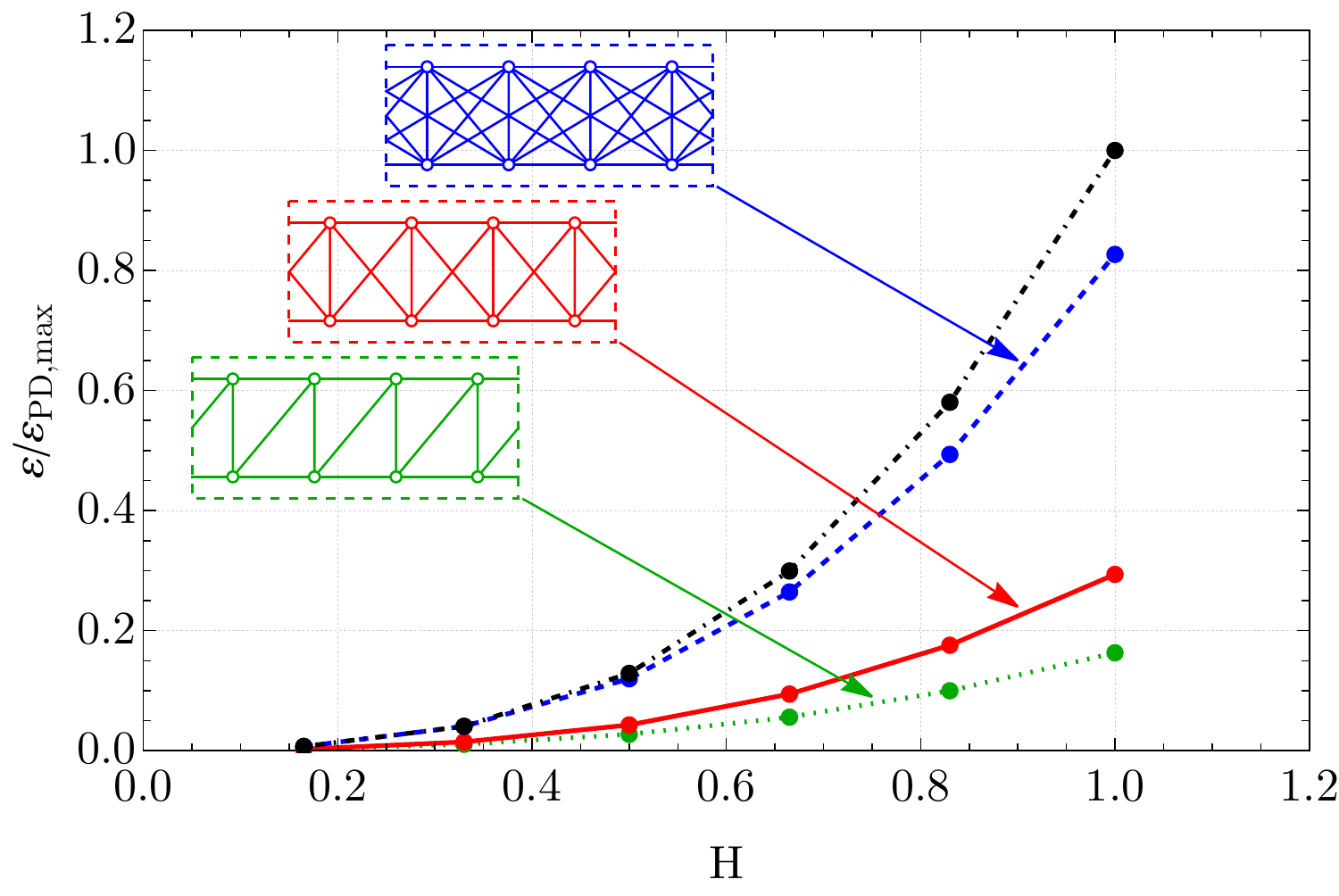}
	\end{center}
	\vspace{-5mm}
\caption{\footnotesize On the left: scaling, as a function of the total height H, of the energy associated with a uniform stretching deformation (purely membrane) for a standard Pratt truss (in green), a hyperstatic truss (in red) and two different patterns of a PD discrete beam; on the right: scaling, as a function of the total height H, of the energy associated to a uniform bending deformation, for a standard Pratt truss (in green), a hyperstatic beam (in red) and two different patterns of a PD discrete beam.}
\label{scalings}
\end{figure}

The structure of bond-based peridynamic constitutive equation (\ref{eq:emmirch}) lends itself to an intuitive and simple physical interpretation. In fact, one can imagine the body under consideration to be uniformly divided into blocks and to substitute each block with a node embodying the mass of that specific block\footnote{In finite-element analysis mesh refinement is a powerful stratagem that allows improving precision and accuracy of the approximated solutions for the problem at hand, in parts of the domain where it is required. Refinement procedures inevitably cause non-uniform discretization of the domain and, in the case of peridynamics, can cause inaccuracy of the solution \cite{CHEN2019}.}. Then, \textit{massless} connections can be introduced between nodes to represent their interactions. These connections must reflect how pairs of particles exert forces onto one another in the PD formulation.\\
In order to do so, one can look at how the energy is stored between pairs of interacting volumes (or areas for a two-dimensional problem), $V_i$ and $V_j$, of the PD continuum model. Taking into account equation (\ref{eq:energyPD}), by means of the mean value theorem for integrals, one has 
\begin{equation}
    \mathcal{E}_{i,j}=\int \limits_{V_i}\int \limits_{V_j} \frac{c}{\sigma}\ \frac{(\boldsymbol{\xi}\cdot \boldsymbol{\eta})^2}{4}\textup{d}V_i\textup{d}V_j =\frac{c}{\sigma^*}(\boldsymbol{\xi}^*\cdot \boldsymbol{\eta}^*)^2 \frac{V_i V_j}{4} = \frac{c}{\sigma^*} \frac{V_i V_j}{4}\ |\boldsymbol{\xi}^*|^2\ |\boldsymbol{\eta}^*|^2\cos^2 \beta\ ,
    \label{eq:discretePD}
\end{equation}
where $\boldsymbol{\xi}^*$ and $\boldsymbol{\eta}^*$ are the relative position vector and relative displacement vector for a pair of points $(\boldsymbol{x}_i^*,\boldsymbol{x}_j^*)$, belonging to the volumes $V_{i}$ and $V_{j}$, for which the theorem holds, and $\beta$ is the angle between them. In the limit of very small volumes $V_{i}$ and $V_{j}$ it is legitimate to assume the $\boldsymbol{\xi}$ and $\boldsymbol{\eta}$ functions to be constant between the volumes of interest and approximate the $(\boldsymbol{x}_i^*,\boldsymbol{x}_j^*)$ with the mid-points of each volume.\\
The right term of equation (\ref{eq:discretePD}) resembles the energy of a truss tilted by the angle between $\boldsymbol{\xi}$ and a horizontal axis \textbf{e}$_1$, and subjected to the edge displacement $|\boldsymbol{\eta}|$.
In this fashion, the stiffness of the bond representing the interaction between volumes would be \begin{equation}
k_{i,j}=\frac{1}{2}\frac{c}{\sigma^*}V_i V_j|\boldsymbol{\xi}^*|^2\ ,
\label{eq:stiffbond}
\end{equation}
where $\boldsymbol{\xi}^*$ is now the relative position vector between mid-points of volumes $V_i$ and $V_j$. Equation (\ref{eq:stiffbond}) implies that the stiffness is proportional to some power $n$ of the distance between particles $k_{i,j}\sim l_{i,j}^n$ (recalling that $\sigma^*=\sigma(|\boldsymbol{\xi}^*|)$).
Performing this type of substitution for all the pairs of nodes transforms the continuum peridynamic body into an intricate reticular beam. The simplest paradigm of structure that can be built by following the PD approximation presented above is depicted in Figure \ref{paradigm}.
The paradigmatic structure characterized by simple kinematics, correctly predicts the hierarchical form of the energy (\ref{eq:wred}) and allows comparing the kinematics of the continuum with something more easily controllable, accompanying the reader in an ideal transition from discrete to continuum for making evident how the above-mentioned nonstandard terms have to naturally appear in the peridynamic plate model.\\
We hereby limit the paradigmatic structure to stretching and bending kinematics which prove to be sufficient for a qualitative interpretation. In particular, the structure is loaded by imposing horizontal displacements at the outer nodes (nodes 1 and 2 in Figure \ref{paradigm}). In this condition, the total energy can easily be obtained analytically.
By denoting $$\overline{\varepsilon}=\frac{u_2+u_1}{4\ l_{14}}\ ,\ \ \overline{\chi}=\frac{2(u_1-u_2)}{l_{12}\ l_{14}},$$ the average stretch and curvature respectively, the total energy amounts to the following expression:
\begin{equation}
    \mathcal{E}_{\textup{discrete}}=\mathrm{E}\ \alpha\ l_{14}^{4+n}\left(\ \textup{H}^2\ \overline{\varepsilon}^2\ \frac{4  \left(1+2\psi^2+\psi^n\right)}{1+\psi^2+\psi^n}+\textup{H}^4\ \overline{\chi}^2\right),
    \label{eq:energy_paradigm} 
\end{equation}
recalling that $l_{12}=$H is the length of truss connecting points 1 and 2 of Fig. \ref{paradigm}, $\psi=\cot \theta$, E is the Young's Modulus of the beams (assumed constant) and $\alpha=c\ V_1^2/2$.
Equation (\ref{eq:energy_paradigm}) represents the equivalent energy of a peridynamic continuum (the horizon is included implicitly since $l_{13}=\delta=l_{14}/\sin{\theta}$). Clearly, membrane energy scales with the square of the thickness H, while bending energy scales with H$^4$ differently from the classical results of local elasticity \cite{STEIGMANN2012,STEIGMANM2020b}. \\
\noindent According to the reduced energy obtained in Section \ref{ch:hierarchical}, the total energy of a plate resembling the shape and loads of the paradigmatic case is recovered by choosing L$=2l_5$, $\delta=l_5/\cos \theta$ (hence $\Psi=(\cos \theta)/2$):
\begin{equation}
    \mathcal{E}_{\textup{PD}} = c\ \sec ^5\theta\ (29+20\cos \theta)\ l_5 ^6 \left(\textup{H}^2\  \overline{\varepsilon}^2\ \frac{108}{29+20\cos 2\theta}+\textup{H}^4\ \overline{\chi}^2\right)\ ,
    \label{eq:paradigmcontinuum}
\end{equation}
which corresponds, at least qualitatively, to (\ref{eq:energy_paradigm}) for $n=2$.
\paragraph{Sensitivity analysis of the paradigmatic structure to the horizon} The total energy of the paradigmatic case has been derived for a fixed value of the horizon, i.e. $\delta=l_5/\cos \theta$, whereas the reduced form of eq. (\ref{eq:wred}) is defined for any value of such parameter. In order to extend the above interpretation analysis using the paradigmatic tool to the general case, a series of numerical analyses using ANSYS APDL has been carried out on several discrete approximations of a peridynamic bond-based continuum each characterized by different horizon sizes.\\
In Figure \ref{scalings} the scaling of three different peridynamic discrete bodies with a horizon ranging from a minimal value (dotted green line) to one with a wider interaction (dot-dashed dark line) is depicted. On the left, the scaling of the membrane energy is found to be quadratic regardless of the horizon size, whereas the bending energy (Figure \ref{scalings}, on the right) scales with the fourth power of the thickness.
A geometric motivation is available for the scaling of the membrane energy: the increase in the number of bonds available when the thickness is doubled, for example, is (roughly) proportional to the square of the number of nodes: $n_{bonds}\sim n_{nodes}(n_{nodes}-1)/2$.

\section{Coupling in the local theory of plates} \label{appendix3}
Eq. (\ref{eq:wred}) is derived from an assumption on the kinematics that is typical of the \textit{first-order shear deformation theory} (FSDT), i.e. Reissner-Mindlin theory, for local plates \cite{REDDY2004,REDDY2014Non}. To further address the previous result about the coupling in eq. (\ref{eq:wred}), we compare the energies for various local plate model, in accordance with the FSDT hypothesis, against the peridynamic result. 
Under the condition of FSDT, small displacements (linear elastic response) and inextensibility in the thickness direction, the specific (per unit area) elastic energy amounts to:
\begin{equation}
    \omega_{\textup{red}}=\frac{\textup{H}}{2}\left((\lambda+2\mu)\mathcal{A}_1'^2+\mu (\mathcal{B}_1+\mathcal{A}_3'^2) \right)+\frac{\textup{H}^3}{24}(\lambda+2\mu)\mathcal{B}_1'^2 \ , 
\end{equation}
\noindent where the $\mathcal{A}_1$, $\mathcal{A}_3$, $\mathcal{B}_1$ and $\mathcal{B}_3$ are to be intended as functions of $x_1$, and $\omega_{\textup{red}}$ is now the reduced energy density of a local plate.\\ 
If one relaxes the inextensibility constraint, meaning the vertical component of the displacement follows the linear approximation in the through-thickness direction, the elastic energy becomes:
\begin{align}
    \omega_{\textup{red}}=&\textup{H}\left\{ (\lambda+2\mu)(\mathcal{A}_1'+\mathcal{B}_3)^2+\mu(\mathcal{A}_3'+\mathcal{B}_1)^2+(\lambda-2\mu)\mathcal{A}_1'\mathcal{B}_3\right\}/2+\\
    &\textup{H}^3\left\{(\lambda+2\mu)\mathcal{B'}_1^2+\mu\mathcal{B'}_3^2 \right\}/24\ .\nonumber
\end{align}
\noindent Lastly, adding a geometric nonlinearity, i.e. small displacements but large/finite rotations hypothesis, to the model leads to:
\begin{align}
    \omega_{\textup{red}}=&\textup{H} \left. \{\mathcal{A'}_3^4 \frac{(\lambda +2 \mu )}{4}+\mathcal{A'}_1^2 (\lambda +2 \mu )+
   \mathcal{A'}_3^2 \left(\mathcal{A}_1' (\lambda +2 \mu )+\mu +\lambda  \mathcal{B}_3\right)+%
    \right.\\ \nonumber
    &\ \ \  \left.2 \lambda  \mathcal{B}_3
   \mathcal{A}_1'+2 \mu  \mathcal{B}_1 \mathcal{A}_3'+ \mathcal{B}_3^2 (\lambda +2 \mu )+ \mu  \mathcal{B}_1^2 \right. \}/2+\\ \nonumber 
    &\textup{H}^3 \left\{40 \mathcal{A}_3' (\lambda +2 \mu ) \mathcal{B}_3'
   \mathcal{B}_1'+ \mathcal{B'}_3^2 \left(\left(3 \mathcal{A'}_3^2+2 \mathcal{A}_1'\right) (\lambda +2 \mu )+2
   \mu +2 \lambda  \mathcal{B}_3\right)+%
    \right. \\ \nonumber
    & \ \ \ \left.  80 (\lambda +2 \mu ) \mathcal{B'}_1^2\right\}/48 \\ \nonumber
    & \textup{H}^5  \left\{(\lambda +2 \mu )\mathcal{B'}_3^4\right\}/640\ \ .\nonumber
\end{align}
From the Euler-Lagrange equations that the previous type of elastic energies can generate, one can see how only the last nonlinear model accounts for the coupling of stretching and bending. The reason for this lies in the nonlinear relation between the deformations and the displacement field. The intrinsic microstructure of a peridynamic continuum naturally accounts for a similar effect, since the finite distances between particles makes it so that rotational effect can be induced by the intricate connections even under axial loading and net null moment.

\bibliographystyle{elsarticle-num}
\bibliography{reference}
\end{document}